\documentclass[11pt,a4paper]{article}
\usepackage{amsmath}
\usepackage{graphics}

\newcommand{\hepth}[1]{{\tt hep-th/#1}}

\newcommand{\nn}{\nonumber}

\newcommand{\p}{\vspace{6pt}\noindent}
\newcommand{\jump}{\vspace{2pt}}
\newcommand{\jumpup}{\vspace{-6pt}}
\advance\textheight by \topskip
\makeatletter

\@addtoreset{equation}{section}
\def\section{\@startsection {section}{1}{\z@}{-8.5ex plus -1ex minus
 -.2ex}{3.3ex plus .2ex}{\large\bf}}
\def\subsection{\@startsection{subsection}{2}{\z@}{-3.25ex plus
 -1ex minus -.2ex}{1.5ex plus .2ex}{\bf}}
\def\subsubsection{\@startsection{subsubsection}{3}{\z@}{-3.25ex plus%
 -1ex minus -.2ex}{1.5ex plus .2ex}{\sl}}

\begin{document}
\begin{titlepage}
\vspace*{-2cm}
\begin{flushright}
YITP-05-29
\end{flushright}

\vspace{0.3cm}

\begin{center}
{\Large {\bf }}
\vspace{1cm}
{\Large {\bf Some aspects of jump-defects in the quantum sine-Gordon model}}\\
\vspace{1cm} {\large P.\ Bowcock\,\footnote {\noindent E-mail: {\tt
peter.bowcock@durham.ac.uk}}, E.\ Corrigan\footnote{\noindent
E-mail: {\tt ec9@york.ac.uk}} and
C.\ Zambon\footnote{\noindent E-mail: {\tt zambon@yukawa.kyoto-u.ac.jp}} }\\
\vspace{0.3cm} {${}^a$}\,\em Department of Mathematical
Sciences\\ University of Durham\\ Durham DH1 3LE, U.K.\\
\vspace{0.3cm} {${}^{b,\,c}\,$\em\it Yukawa Institute for Theoretical Physics\\
University of Kyoto\\
Kyoto 606-8502, Japan\\
\vspace{0.3cm}{${}^b$}\,Department of Mathematics \\ University of York\\
York YO10 5DD, U.K. }\\

\vspace{1cm}
{\bf{ABSTRACT}}

\end{center}

\begin{quote}
The classical sine-Gordon model permits integrable discontinuities, or jump-defects, where
the conditions
relating the fields on either side of a defect are  B\"acklund transformations frozen at
the defect location. The purpose of this article is to explore the extent to which this idea
may be extended to the quantum sine-Gordon model and how the striking features of the classical model
may translate to the quantum version. Assuming a positive defect parameter there are two types of
defect. One type, carrying even charge, is stable, but the other type, carrying odd charge, is
unstable and may be considered as a resonant bound state of a soliton and a stable defect. The
scattering of solitons with defects is considered in detail, as is the scattering of breathers,
and in all cases the jump-defect is purely transmitting.
One surprising discovery concerns the lightest breather. Its transmission factor is independent
of the bulk coupling - a property susceptible to a perturbative check, but not shared with any
of the other breathers. It is argued that classical jump-defects can move and some comments
are made concerning their quantum scattering matrix.
\end{quote}

\vfill
\end{titlepage}
\section{Introduction}
\jumpup
Recently, there has been some interest in the question of whether or not defects (or impurities) of
various kinds might be integrable and, if so, what kind of properties they might have.
Already ten years ago it was pointed out by Delfino, Mussardo and Simonetti \cite{Delf94} that the
standard ideas of factorisation with a bulk, Lorentz invariant, S-matrix are incompatible
(in most cases) with both
reflection and transmission at a defect (the exceptions occurring  when the S-matrix is
$\pm 1$, or possibly when the defect has internal degrees of freedom \cite{Fring02}).
Also, Konik and LeClair \cite{Konik97} have examined the possibility of having
a purely transmitting defect in the sine-Gordon model, in the sense of investigating
algebraically the
equations satisfied by the quantum transmission matrix. Some of their results
appear to be relevant to the case of interest in this article although their
starting point and emphasis were very different. More recently, some of the standard heuristic
ideas have been questioned and modified within a different scheme developed by Mintchev, Ragoucy and
Sorba \cite{Mintchev02} and applied to various models \cite{Caudrelier04}. Other
ideas  can be
found in \cite{Bajnok} and \cite{Moriconi}. However,
the principal focus of  those works has been to combine reflection and transmission,
whereas the present article will focus exclusively on purely transmitting integrable defects
of a quite particular type.

\p In \cite{bczlandau} it was noted,
in a Lagrangian approach, that a field theory could permit a discontinuity
or `jump-defect'
and yet remain classically integrable. The principal examples of this phenomenon
discovered so far have been very specific. These include free scalar fields, Liouville theory,
the sine-Gordon model,
and certain affine Toda models \cite{bcztoda}, all of which permit  B\"acklund transformations.
Indeed, the defect conditions relating fields evaluated as limits from either side
of the jump-defect turn out to be a B\"acklund transformation frozen at the location of
the defect. This fact is quite striking in view of the importance B\"acklund
transformations have played in
the development of soliton theory. By now there is an extensive literature on
 B\"acklund transformations, and their uses, in a variety of
contexts; see for example \cite{Backlund}. A particular feature of this kind of jump-defect
is precisely that it is purely
transmitting at the classical level, and presumably also at the quantum level. In a sense,
a jump-defect is more severe than the more usual model of an impurity represented by
a delta-function contribution in the equations of motion, since the latter generally
requires the field, though not necessarily its spatial derivatives, to be continuous.
On the other hand, a defect of delta-function type does not generally maintain classical
integrability, although there are a variety of other interesting phenomena involving solitons
associated with it \cite{Goodman2002}. A jump-defect is also simpler because, in the
examples found so far, the classical
systems have no periodic solutions specifically associated with the presence of the defect.
This is in contrast to integrable field theory with a boundary in which boundary breathers
may be found for a suitable range of parameters even in the simplest of models \cite{cd}.

\p It is the purpose of this article to investigate the quantum sine-Gordon field
theory in the presence of a jump-defect and to locate within the quantum framework
the influence of at least some of the striking consequences of the classical model.
Details are provided in \cite{bczlandau} but for convenience, a brief summary
will also be provided in the next section.
Particular attention will be paid to the properties of solitons scattering with the
defect. For reviews of models with solitons in general, including the
sine-Gordon model, see for example \cite{Scott73}; much essential information
concerning the quantum
sine-Gordon
model is to be found in the review by Zamolodchikov and Zamolodchikov \cite{Zam79}.

\section{Classical sine-Gordon with a jump-defect}

\jumpup The sine-Gordon model in the bulk will be taken to be defined by the Lagrangian density
\begin{equation}\label{sGbulk}
    {\cal L}=\frac{1}{2}\left((\partial_t u)^2-(\partial_x u)^2\right) - \frac{m^2}{\beta^2}
    (1-\cos\beta u),
\end{equation}
although for classical considerations it is often convenient to remove the mass parameter $m$ and
the coupling $\beta$ by a rescaling.
A single jump-defect placed at $x=0$ is described by modifying the Lagrangian in the following
manner,
where the field on the left of the defect is denoted by $u $ and the field on the right of it
by $ v $. The full Lagrangian will consist of pieces from the bulk regions ($x<0$ and $x>0$),
together with a delta function contribution at $x=0$. In detail,  the Lagrangian density is given by
\begin{equation}\label{sGdefect}
{\cal L}=\theta(-x){\cal L}_u +\theta(x){\cal L}_v-
\delta(x)\left[\frac{1}{2}\left(u  v _t- v  u _t\right) -{\cal B}(u , v )\right]
\end{equation}
with
\begin{equation}\label{sGdefectpotential}
    {\cal B}=-\frac{2m\sigma}{\beta^2}\cos\beta\left(\frac{u + v }{2}\right)-
    \frac{2m}{\sigma\beta^2}\cos\beta
    \left(\frac{u - v }{2}\right).
\end{equation}
Following from this the (suitably scaled) bulk field equations and defect conditions are:
\begin{eqnarray}\label{sGdefectequations}
\nn &&x<0:\quad\partial^2u =-\sin u ,\quad\\
\nn &&x>0:\quad\partial^2 v =-\sin v ,\\
&&x=0:\quad \partial_xu -\partial_t v =
-\sigma\sin\left(\frac{u + v }{2}\right)-\frac{1}{\sigma}
\sin\left(\frac{u - v }{2}\right)\nn\\
&&\phantom{x=0:}\quad\ \partial_x v -\partial_tu =
\phantom{-}\sigma\sin\left(\frac{u + v }{2}\right)-\frac{1}{\sigma}
\sin\left(\frac{u - v }{2}\right)\label{condition}.
\end {eqnarray}
The term containing first order time derivatives in the defect Lagrangian  is required by
integrability and has interesting consequences. Clearly, it is not
invariant under reversing the sense of time, nor is it invariant under
the bulk symmetries $u\rightarrow u+2a\pi/\beta, \ v\rightarrow v+2b\pi/\beta$, where $a$ and $b$
are integers; it is not even invariant under the reflections $u\rightarrow -u$ or
$v\rightarrow -v$. However, it
is invariant under certain combinations, such as reflecting both fields simultaneously
or reflecting one of them and reversing the sense of time.
Under the bulk symmetries, which translate among the different bulk ground states, the defect
term changes by a total time derivative (note that the defect potential itself
is invariant under these transformations only if $a\pm b$ is even).
However, that is insufficient for invariance under all
circumstances because how the action changes will depend upon boundary conditions -
in this sense it is reminiscent of a
one-dimensional `Chern-Simons' term - and these may vary in the presence of scattering
configurations of fields. Additional comments concerning
this aspect in the quantum field theory will be made at the end of section (3).

\p
The Lagrangian does not violate time translation and,  provided a contribution
from the defect, involving the fields evaluated at
the defect, is included, the total energy, $${\cal E}={E}(u )+{ E}( v )+{\cal B},$$ will be conserved.
Moreover, setting
$$\frac{\partial{\cal U}}{\partial u }=\frac{\partial{\cal B}}{\partial v },\quad
\frac{\partial{\cal U}}{\partial v }=\frac{\partial{\cal B}}{\partial u },$$
so that,
$${\cal U}=-2\sigma\cos\left(\frac{u + v }{2}\right)+\frac{2}{\sigma}\cos
    \left(\frac{u - v }{2}\right),$$
one finds that the total momentum, $${\cal P}={P}(u )+{P}( v )+{\cal U},$$ is
also conserved \cite{bczlandau}.
Hence the fields on either side of $x=0$ can exchange both energy and momentum
with the defect.  In fact, the conditions \eqref{sGdefect} are immediately recognisable
as a B\"acklund transformation frozen at $x=0$ and it would be attractive to have a
specific physical mechanism to implement it. It is worth emphasising that the defect potential has
twice the period of the bulk potentials, a feature it shares with the potentials representing
integrable boundary conditions \cite{Ghosh94a}, which restrict the sine-Gordon model to
a half-line without destroying its integrability. It has already been noted that the defect
Lagrangian is not invariant under time-reversal or parity. Rather there is in each case an
extended symmetry involving the fields and the parameter $\sigma$. In the case of parity
this will be further remarked upon below.

\p In the sine-Gordon model a single soliton \cite{Scott73} may be described conveniently by
\begin{equation}\label{sGsoliton}
    e^{iu /2}=\frac{1+iE}{1-iE}, \quad E=e^{a x +b t +c},
\end{equation}
where $a, \ b,\ e^c$ are all real, with $a^2-b^2=1$, and it is useful
to parameterize $a$ and $b$
in terms of rapidity, setting $a=\cosh\theta,\ b=-\sinh\theta$. With $\theta>0$,
the soliton is moving along the $x$-axis in a positive direction. An anti-soliton (having the same
velocity and location) is
obtained by the replacement $E\rightarrow -E$ (or equivalently $c\rightarrow c+i\pi$).
The mass of the soliton  is 8 in the units for
which the fields $u$ or $v$ have unit mass parameter. [If the coupling and mass scale are reinserted,
the mass parameter of the fields will be $m$ and that of the soliton will be $8m/\beta^2$.]

\p
Supposing  a soliton given by \eqref{sGsoliton}, moving in a positive sense along the $x$-axis,
encounters the defect, then a similar, but delayed,  soliton emerges given by
\begin{equation}\label{sGsolitonout}
    e^{i v /2}=\frac{1+izE}{1-izE}, \quad E=e^{a x +b t +c},
\end{equation}
where $z$ represents the delay \cite{bczlandau}. Using the defect conditions,
\begin{equation}\label{delay}
    z=\frac{e^{-\theta}+\sigma}{e^{-\theta}-\sigma},
\end{equation}
and there will be a variety of possible consequences according to the choice of $\sigma$.
Setting $\sigma=e^{-\eta}$, the expression for $z$ can be written alternatively as
\begin{equation}\label{}
    \nn z= \coth\left(\frac{\eta-\theta}{2}\right).
\end{equation}

\p
If $\theta >0$ and if $\eta <0$ it is clear $z<0$, implying that the incoming soliton
always converts to an anti-soliton, although for large $\theta$ it will be delayed very little.
On the other hand, for $\eta>0$ there are several possibilities: if $\theta < \eta$ the
soliton will be delayed but its character remains unchanged; if $\theta >\eta$ the soliton
flips to an anti-soliton; but if $\theta=\eta$ the incoming soliton is infinitely delayed
and therefore never emerges from the defect. In effect, a soliton, which in the distant
past interpolated between $0$ and $2\pi$, is replaced in
the far future by the static solution $u=0,\ v=2\pi$. It is easy to check that the
latter satisfies
the defect conditions by itself and stores at the defect location precisely the
energy and momentum originally carried by the soliton. A soliton travelling in the opposite
direction
will be affected in a similar variety of ways. Several solitons passing the same defect
are each delayed by a similar factor. In effect, the multi-soliton solution $u$ constructed
by assembling, using Hirota's method \cite{Hirota}, a set of exponential factors
$$E_k=e^{a_kx+b_kt+c_k},$$
is replaced by a similar solution $v$ in which each exponential factor is multiplied by
 the appropriate factor $z_k$, given by \eqref{delay} with $\theta$ replaced by $\theta_k$.

 \p There might be many defects placed at different locations, each with its own parameter,
 and each affecting a passing soliton independently of all the others. In particular,
 a soliton passing two defects, which are
separated spatially
but otherwise identical, always retains its character (since if it is flipped by one it
must be flipped by the other). However, it will be delayed by
the combined factor
$$z=\coth^2\left(\frac{\eta-\theta}{2}\right).$$
Interestingly, this is precisely the delay it would have experienced had it been overtaken
(or been overtaken by) a soliton of rapidity $\eta$ \cite{Scott73}. Moreover, a current of
such solitons
would build up
positive and negative topological charges on the two defects - indicating that a pair of
similar defects might behave like a store of topological charge. In other words, a pair
of similar defects might be thought of as an analogue
of a capacitor in an electrical circuit.

\p The behaviour of solitons in the presence
of jump-defects suggests that if
mathematical defects such as these can be found within actual physical systems then
using them to control solitons might lead to
technological applications, perhaps along the lines suggested in \cite{cz}. They
also suggest that the jump-defect behaves, in a sense, as though it were `half' a soliton.
Even the energy or momentum naturally associated with a defect via the expressions for the total
energy and momentum relate to an object of mass $4m/\beta^2$ rather than one of mass $8m
/\beta^2$.

\p
Since any number of jump-defects can co-exist, each
influencing the progress of a soliton independently of all the others, and since they
are all at rest, it is clear the defects do not exert any nett long-range influence on
each other. In this respect, the defects are quite different to classical sine-Gordon
solitons
since multi-soliton solutions with each soliton at rest at an arbitrarily chosen
location cannot exist. Famously,  there are extended structures in three spatial
dimensions which may be assembled to create multi-particle-like stationary solutions
to their equations of motion. Yang-Mills-Higgs BPS-monopoles provide a prime example of
this phenomenon (for a
recent review, see \cite{Manton04}): classical solutions can be constructed with arbitrary
numbers of like-charged magnetic monopoles balancing at rest despite the existence of long-range
forces between them. In this case, there are two kinds of long range force (the
electromagnetic force and the Higgs force), exactly
cancelling out.  There is no reason, in principle, why the defects
should not move, and some comments concerning that possibility will be made
in section(6). If they are able to move with different speeds then inevitably they must
scatter and the most interesting question concerns the nature of this scattering and the
nature of the associated short-range interaction. A further question will be whether
the defects themselves are describable by a quantum field theory.

\p
For future reference, it is also instructive to calculate the transmission factor through
the defect in the situation where the equations \eqref{sGdefect} are linearized. If the
linear perturbation is a perturbation of the static solution $u=2n\pi, v=2m\pi$, then
it will be denoted $T_{\rm even}$ or $T_{\rm odd}$ according to whether $n-m$ is even or odd.
It is straightforward in either case to show there is no reflection and $u$ and $v$ will
have the form
\begin{equation}\label{Tbreatherclassical}
u=e^{-i\omega t+ik x}, \quad x<0; \quad v=Te^{-i\omega t+ik x}, \quad x>0
\end{equation}
with\begin{equation}\label{classicalT}
    T_{\rm even}(\theta, \eta)=-i\, \frac{\sinh\left(\frac{\theta-\eta}{2}-\frac{i\pi}{4}\right)}
    {\sinh\left(\frac{\theta-\eta}{2}+\frac{i\pi}{4}\right)}=\bar T_{\rm odd}(\theta,\eta).
\end{equation}
In the limit $\eta\rightarrow\infty$, (or $\sigma\rightarrow 0$), the transmission factor
tends to unity, as it should since continuity of the fields is restored and the jump
disappears. Notice that the transmission factors are unitary but they do not satisfy
$T(\theta)T(-\theta)
=1$. This is a consequence of the behaviour of the defect conditions under parity or
time reversal, neither of which is an invariance of \eqref{sGdefect}.
For example, a parity transformation interchanges $u$ and $v$ but
needs to be accompanied by the change of sign of one of the fields $u$ or $v$ and a
replacement of $\sigma$ by $1/\sigma$, or equivalently, $\eta \rightarrow -\eta$.
Thus, defining $T_P(\theta,\eta)=-T(-\theta, -\eta)$ one finds
\begin{equation}\label{}
    T(\theta)T_P(\theta)=1.
\end{equation}
However, for a specific $\sigma$ both parity and time-reversal are explicitly broken.

\p
In the quantum field theory it is expected that the bound states, or `breathers'
will be transmitted through a defect and suffer a change of phase whose classical limit
should be one of $T_{\rm even}$ or $T_{\rm odd}$, depending on the precise circumstances.

\p
A couple of other observations are in order.
It has already been pointed out that the defect conditions \eqref{sGdefect} have the
form of a B\"acklund transformation
with the spatial derivatives fixed at the location of the defect. In the bulk, a
B\"acklund transformation
generates solitons in the sense that if one of the fields is taken to be zero and the
equation for the other is integrated, the result is typically a one-soliton solution.
Similarly, if the first is taken to be a one-soliton, integrating for the second will
give a two-soliton, and so on \cite{Scott73}. The question is: how does this behaviour fit in with
the defect conditions given above?
Clearly, since the defect conditions are a frozen B\"acklund transformation, there is no
question of integrating the equations for one field given the other.
Up to this point the conditions have been used to demonstrate how a single soliton approaching the
defect will be delayed on passing through, or possibly altered more drastically. However,
the possibility of two solitons emerging has not been considered.

\p
Consider the possibility that \eqref{sGdefect} allows a single soliton described
by $u$ to approach the defect, and a two-soliton, described by $v$ to emerge from
it. It is not difficult to check that for this to happen one of the emerging solitons
must be a delayed version of the original one and the new soliton has parameters related
to the defect parameter $\sigma$ (although there is no information that would fix
its `position').
If this is possible then as $t\rightarrow\infty$, $E(v)$ must have a contribution from
both solitons, whereas as $t\rightarrow -\infty$, $E(u)$ has a contribution merely from
one of them. To balance the energy, the additional contribution must come from the difference of the
energies initially and finally stored in the defect. Supposing $\sigma >0$, and initially
there is no discontinuity at the defect (that is, $u(0,-\infty)=2\pi=v(0,-\infty))$,
then the defect contribution to the energy \eqref{sGdefectpotential} will be negative and equal to
\begin{equation}\label{}
    -2\left(\sigma+\frac{1}{\sigma}\right)=-4\cosh \eta.
\end{equation}
On the other hand, if subsequently there are two solitons on the right, the defect must
end up with a $2\pi$ discontinuity, meaning the energy stored there must have {\it increased}
since it will ultimately have to be
\begin{equation}\label{}
2\left(\sigma+\frac{1}{\sigma}\right)=4\cosh\eta.
\end{equation}
Since the additional soliton also contributes positively to the energy the overall energy
conservation is violated. So, this situation cannot actually occur. The only possibility
in these circumstances is for the approaching soliton to pass through - albeit with a delay (or
conversion to an anti-soliton). Hence, no energy is extracted from the defect and the
final configuration of the fields at the defect is $u(0,\infty)=0=v(0,\infty)$.
Another possible starting configuration has a $2\pi$ discontinuity at the defect, meaning a
store of positive energy which is exactly right to allow a new soliton (or anti-soliton)
to emerge in the final state, leaving a $4\pi$ (or $0$) discontinuity behind. The new soliton
will have rapidity $\theta =\eta$ and energy $E=8\cosh\eta$. However, the classical
system provides no information as to the relative position, or character, of the additional
soliton.

\p
Since $\sigma$ is a free parameter, it is also possible to take $\sigma<0$. Then
the energy stored in the defect would be positive to begin with provided the initial
discontinuity was zero modulo $2\pi$. Then, in principle, a soliton could emerge making use of
that energy leaving a discontinuity behind.

\p
Thus it appears defects might produce solitons as well as absorbing
them. However, since there is no information concerning the  location (or time-origin)
of the additional soliton the situation bears a resemblance to an excited atom.
Thought of classically, there is no information to indicate the time of decay. Instead, quantum
mechanics is needed to supply a probability of decay within a specified time. This analogy
suggests the quantum story of defects of the kind considered here could be
considerably more interesting and motivates a search for the quantum analogues of the
transmission matrices.

\p
To date, the discussion has been entirely theoretical
and it would be even more interesting to find a physical situation
in which B\"acklund
transformations
play a substantive role, rather than being a mathematical, solution-generating, device.
In such a situation it would be expected that the effects described briefly above
should be manifest and amenable to observation.

\section{Transmission matrices}
\jumpup
Rather than depending upon the previous literature \cite{Konik97} for the transmission
matrices they will be derived afresh from first principles, guided by the classical features
of the jump-defect. In any case, there will be some significant differences with respect to
earlier work on this topic.

\p
On the basis of the classical scattering of a soliton from the jump-defect it
is expected the defect will be purely transmitting and able to store topological charge.
Moreover, topological charge may be added or removed in steps of two.
The classical picture suggests
there will be two types
of transmission matrix depending on whether the defect is carrying an even or odd charge.
If $\sigma >0$, only the even transmission matrix is expected to be unitary
since the `vacuum', or least energy configuration of a defect with a positive
parameter $\sigma$, must have even (positive, negative, or zero) topological charge
and cannot decay.
 Thus, there should be a
transmission matrix $$^{\rm e}T_{a\alpha}^{b\beta}(\theta),$$ regarded as describing the
transmission from the region
$x<0$ to the region $x>0$, where $a$ and $b$ may be $+$ for a soliton or $-$ for an anti-soliton,
and the labels $\alpha$ and $\beta$ are even integers (positive, negative or zero). These
transmission matrices should satisfy (for real rapidity $\theta$),
\begin{equation}\label{Tevenunitarity}
    ^{\rm e}T(\theta)^{\rm e}T^\dagger(\theta)=1.
\end{equation}
More precisely, the transmission matrices relate states of the system
in the far future to those in the far past and the state of the system is labelled by the
soliton energy-momentum, or rapidity, and its character (topological charge) together with the state
of the defect  labelled by the topological charge accumulated on it.
On the other hand, because of the properties of the Lagrangian under a parity transformation
the transmission matrix for a soliton moving in the opposite direction will be different.

\p
The transmission matrix will depend also on the
defect parameter and the bulk coupling as well as rapidity, and satisfy a number of other relations
to be detailed below.

\p
The soliton-soliton bulk scattering matrix is taken to be the standard one \cite{Zam79}, given by
\begin{equation}\label{Smatrix}
    S_{kl}^{mn}(\Theta)=\rho(\Theta)\left(%
\begin{array}{cccc}
  a(\Theta)& 0 & 0 & 0\\
  0 & c(\Theta)& b(\Theta) & 0 \\
  0 & b(\Theta) & c(\Theta) & 0\\
  0 & 0 & 0 & a(\Theta)\\
\end{array}%
\right),
\end{equation}
where $k,l$ label the incoming particles and $m,n$ label the outgoing particles
in a two-body scattering process, with the particles labelled $k,n$ having rapidity
$\theta_1$, and the particles labelled $l,m$ having rapidity $\theta_2$. The various pieces
of the matrix are defined by
\begin{equation}
\quad a(\Theta)=\frac{qx_2}{x_1}-\frac{x_1}{qx_2},\quad b(\Theta)=
\frac{x_1}{x_2}-\frac{x_2}{x_1},\quad c(\Theta)=q-\frac{1}{q},
\end{equation}
with
\begin{equation}\label{}
    \Theta=\theta_1-\theta_2,\quad q=e^{i\pi\gamma},\quad x_p=e^{\gamma\theta_p}.
\end{equation}
In this notation the crossing property of the S-matrix is represented by
\begin{equation}\label{Scrossing}
    S_{k\,l}^{m\,n}(i\pi-\Theta)=S_{k\,\, -m}^{-l\, \,n}(\Theta),
\end{equation}
with the diagonal elements $S_{+-}^{+-}(\Theta)$ and $S_{-+}^{-+}(\Theta)$ crossing into
themselves.
The overall factor $\rho(\Theta)$ will be needed later and is given by:
\begin{equation}\label{Smatrixrho}
    \rho(\Theta)=\frac{\Gamma(1+i\gamma\Theta/\pi)\Gamma(1-\gamma -i\gamma\Theta/\pi)}{2\pi i}
    \prod_{k=1}^\infty\, R_k(\Theta)R_k(i\pi -\Theta),
\end{equation}
where
\begin{equation}\label{}
    R_k(\Theta)=\frac{\Gamma(2k\gamma +i\gamma\Theta/\pi)\Gamma(1+2k\gamma +i\gamma\Theta/\pi)}
    {\Gamma((2k+1)\gamma +i\gamma\Theta/\pi)\Gamma(1+(2k-1)\gamma +i\gamma\Theta/\pi)}.
\end{equation}

\p Note, the conventions adopted by Konik and LeClair \cite{Konik97} have been used. Therefore,
in particular, the coupling
$\gamma$ in terms of the Lagrangian coupling $\beta$ (with $\hbar=1$ and the conventions
indicated by
\eqref{sGbulk})
is defined by
\begin{equation}\label{gamma}
    \frac{1}{\gamma}=\frac{\beta^2}{8\pi-\beta^2}.
\end{equation}

\p
Since the defect is purely transmitting, the usual heuristic arguments based on
factorisability and bulk integrability \cite{Delf94} would require
\begin{equation}\label{STT}
    S_{kl}^{mn}(\Theta)\, ^{\rm e}T_{n\alpha}^{t\beta}(\theta_1)\,
    ^{\rm e}T_{m\beta}^{s\gamma}(\theta_2)
    =\, ^{\rm e}T_{l\alpha}^{n\beta}(\theta_2)\, ^{\rm e}T_{k\beta}^{m\gamma}(\theta_1)\,
    S_{mn}^{st}(\Theta),
\end{equation}
and this seems to be the most appropriate assumption to make in the present context.
There is also a transmission matrix representing transmission through the defect
of a soliton moving from right to left. However, this will be determined in terms
of $^{\rm e}T$ and part of the purpose in this section  is to develop a set of
criteria constraining $^{\rm e}T$ itself without reference to transmission from
right to left. This means that the important crossing properties of  $^{\rm e}T$ will
not be part of the initial story. This may appear to be an unconventional way to proceed
but it is useful to disentangle those aspects of the scheme which are purely algebraic,
or depend upon conjectured general principles, from those which depend upon the special
nature of the Lagrangian model defined via \eqref{sGbulk} and \eqref{sGdefect}.

\p
The use of a bulk scattering matrix depending only on the
rapidity difference might be questioned in a situation which appears manifestly to break Lorentz
invariance. However, to decide that question will require a full discussion of
moving defects going beyond the scope of the present paper. For the time
being \eqref{STT} will be used as a working hypothesis to be abandoned,
if necessary,  when the details emerge of a much fuller picture. This will
include the scattering of defects themselves. One indication that the defect
will itself behave like a particle lies in the results obtained below where the transmission
matrices definitely depend upon the difference of $\theta$, the rapidity of a soliton,
and $\eta$, which in view of the classical defect conditions being a frozen B\"acklund
transformation has the character of a rapidity. Indeed, a Lorentz transformation
on the system will be compensated by a change of the parameter $\sigma$ and a shifting
in the location of the defect. The scattering of defects is likely to be entirely consistent
classically owing to Bianchi's celebrated theorem of permutability for B\"acklund
transformations \cite{Backlund}. Further remarks on this will be made in section (6)
where a start on  the construction of the quantum scattering matrix will also be given.

\p Equations \eqref{STT} may have several solutions and the purpose of this article is to
find among these one which matches qualitatively the jump-defect situation and to find ways
to accumulate evidence for it.
For some purposes it is convenient to change notation slightly and write
\begin{equation}\label{blockT}
    ^{\rm e}T=\left(%
\begin{array}{cc}
  T_+^+ & T_+^- \\
  T_-^+ & T_-^- \\
\end{array}%
\right)\equiv
\left(%
\begin{array}{cc}
  A & B\\
  C & D \\
\end{array}%
\right),
\end{equation}
where the block matrix entries are labelled by the topological charge of the defect (and
are therefore infinite dimensional).

\p
For a first examination, it is useful to make use of the topological charge conservation to note that
$A,D$ are diagonal while $B,C$ are slightly off-diagonal:
\begin{equation}\label{ABCD}
    A_\alpha^\beta=a_\alpha\delta_\alpha^\beta, \quad D_\alpha^\beta=
    d_\alpha\delta_\alpha^\beta,\quad
    B_\alpha^\beta=b_\alpha\delta_\alpha^{\beta-2},\quad C_\alpha^\beta=
    c_\alpha\delta_\alpha^{\beta+2}.
\end{equation}
Then, the transmission relations reduce to a number of equations among the matrices $A,B,C,D$
and the entries of the $S$-matrix, which fall into three types. The first is a set of four two-term
relations (for the purposes of these the subscripts or superscripts `1' and `2' refer to the
rapidities of the incoming particles):
\begin{equation}\label{twoterm}
    A_1A_2=A_2A_1,\ D_1D_2=D_2D_1,\ B_1B_2=B_2B_1,\ C_1C_2=C_2C_1.
\end{equation}
The first two  are automatically satisfied since $A$ and $D$ are diagonal but the other pair provide
genuine constraints. Consider the third: in terms of components one finds
\begin{equation}\label{BBequation}
 b_\alpha^1b_{\alpha+2}^2=b_\alpha^2 b_{\alpha+2}^1\quad \hbox{or}\quad \frac{b_{\alpha+2}^1}
{b_{\alpha}^1}=\frac{b_{\alpha+2}^2}{b_{\alpha}^2},
\end{equation}
and the second of these implies that neither side can depend on the rapidity, implying
\begin{equation}\label{BB}
    b_\alpha(\theta)=\sigma^{\alpha}\, b_0(\theta),
\end{equation}
where $\sigma$
has no dependence on rapidity. In fact, there are two such solutions according to whether
 $\alpha$ is selected to be even or odd. Similarly, the solution for the components of $C$ are:
\begin{equation}\label{CC}
c_\alpha(\theta)=\tau^{\alpha}\, c_0(\theta),
\end{equation}
where $\tau$ has no dependence on rapidity. A second group of four equations has the form
\begin{eqnarray}\label{fourterm}
\nn  b( B_1C_2-C_2B_1) &=& c(D_2A_1-D_1A_2)  \\
\nn  b( B_2C_1-C_1B_2)&=&c(A_1D_2-A_2D_1) \\
\nn  b(A_1D_2-D_2A_1)&=& c(C_2B_1-C_1B_2)\\
  b(A_2D_1-D_1A_2)&=& c(B_1C_2-B_2C_1).
\end{eqnarray}
Since $A,D$ are diagonal, the last pair relate $B$ and $C$:
\begin{equation}\label{BC}
    \frac{c_{\alpha+2}^1}{b_\alpha^1}=\frac{c_{\alpha+2}^2}{b_\alpha^2},
\end{equation}
implying that neither side depends on $\theta$. Hence, to ensure this works for
any $\alpha$ requires a constraint. Introducing a
$\theta-$independent constant $\mu$ this is
\begin{equation}\label{}
    c_0(\theta)=\mu\, b_0(\theta).
\end{equation}
Using \eqref{BC} the difference of the first two of \eqref{fourterm} becomes an identity
leaving one equation to investigate later. The third set of eight equations have the
form,
\begin{eqnarray}\label{threeterm}
\nn aA_1B_2\ =\ bB_2A_1 +cA_2B_1, && \quad aB_1A_2\ =\ bA_2B_1 +cB_2A_1,\\
\nn aA_2C_1\ =\ bC_1A_2 +cA_1C_2, && \quad aC_2A_1\ =\ bA_1C_2 +cC_1A_2,\\
\nn aD_2B_1\ =\ bB_1D_2 +cD_1B_2, && \quad aB_2D_1\ =\ bD_1B_2 +cB_1D_2,\\
aD_1C_2\ =\ bC_2D_1 +cD_2C_1, && \quad aC_1D_2\ =\ bD_2C_1 +cC_2D_1.
\end{eqnarray}
Using \eqref{ABCD}, the first pair of these can be combined to give
\begin{equation}\label{AA}
    \left(a\, \frac{a_\alpha^1}{a_{\alpha+2}^1}-b\right)\left(a\,
    \frac{a_{\alpha+2}^2}{a_\alpha^2}-b\right)=c^2,
\end{equation}
which, on using the definitions of $a,b,c$, implies
\begin{equation}\label{alternatives}
\frac{a_{\alpha+2}}{a_\alpha}=-q \quad \hbox{or} \quad -1/q.
\end{equation}
For either of the two choices there will be a solution for $A$. For example,
choosing the first and setting $-q=Q^2$ the solution has the form,
\begin{equation}\label{evenA}
    a_\alpha(\theta)=Q^{\alpha}a_0(\theta).
\end{equation}
Actually, since $\alpha$ may be a positive or negative integer both possibilities
given in \eqref{alternatives} are covered by \eqref{evenA}.
Inserting this in the first of the first pair in \eqref{threeterm} reveals
\begin{equation}\label{AB}
   x_2 \, \frac{a_0^2}{b_0^2}= x_1\, \frac{a_0^1}{b_0^1}
\end{equation}
implying that
\begin{equation}\label{}
    b_0(\theta)=\lambda\, x_\theta\, a_0(\theta),
\end{equation}
with $\lambda$ independent of rapidity. Hence also (slightly redefining $\mu$),
\begin{equation}\label{}
c_0(\theta)=\mu\, x_\theta\, a_0(\theta).
\end{equation}
The second pair of equations involving $A,C$ is
now an identity. The two equations of the third pair when combined reveal a similar
relation to \eqref{AA} and therefore
$D$ can also be written in the form
\begin{equation}\label{}
    d_\alpha(\theta)=Q^{\, \epsilon\alpha}d_0(\theta),\quad
    \epsilon =\pm 1.
\end{equation}
In addition, the first of
the third pair also reveals,
\begin{equation}\label{}
    d_0(\theta)=\nu x_\theta^{1+\epsilon}a_0(\theta),
    \quad \epsilon =\pm 1.
\end{equation}
With these expressions for $A,B,C,D$, each of \eqref{threeterm} is an identity. Finally, it
is necessary to return to the first of the four-term relations \eqref{fourterm}. It now implies
\begin{equation}\label{}
b\, x_1x_2 (\sigma\tau)^\alpha\mu\lambda\left(\tau^2-\frac{1}{\sigma^2}\right)
=c \, Q^{(1+\epsilon)\alpha} \nu\left(x_2^{1+\epsilon}-x_1^{1+\epsilon}\right).
\end{equation}
If $\epsilon=-1$ then $\tau^2=1/\sigma^2$ is required but
there are no other constraints on the coefficients.
On the other hand, if $\epsilon=1$ one requires,
\begin{equation}\label{tausigmaconstraints}
    \sigma\tau =Q^2, \qquad \nu = -\, \frac{q\lambda\mu}{\sigma^2}\, .
\end{equation}
In view of \eqref{tausigmaconstraints} it would be convenient in this case to put
$\sigma=\rho Q,\ \tau =Q/\rho$, with $\rho$ a free parameter, and then
\begin{equation}\label{}
    \nu= \frac{\lambda \mu}{\rho^2}.
\end{equation}

\p
At this point it is convenient to summarize the possibilities  as follows:
\begin{eqnarray}\label{summary}
\nn    &&a_\alpha=Q^{\, \alpha}A(\theta),\quad d_\alpha=\nu Q^{\, \epsilon\alpha}
x_\theta^{1+\epsilon}A(\theta),\quad
    \epsilon=\pm 1\\
 \nn   && b_\alpha=\lambda \sigma^\alpha x_\theta A(\theta),\quad c_\alpha=\mu
 \sigma^{-\alpha}x_\theta
    A(\theta), \quad \hbox{if}\ \epsilon=-1\\
 && b_\alpha=\lambda (\rho Q)^\alpha x_\theta A(\theta),\quad c_\alpha=\mu (Q/\rho)^\alpha
x_\theta
    A(\theta), \quad  \nu=\mu\lambda/\rho^2 \quad \hbox{if}\ \epsilon=+1.
\end{eqnarray}
The quantities $\mu, \nu, \lambda $ and $A(\theta)$ will differ between the
two alternative choices of $\epsilon$. Notice that the whole process, starting from \eqref{STT}
could be repeated
assuming the defect labels on the $T$-matrix were odd since it transpires the even and
odd solutions never mix via \eqref{STT}. On the other hand, for the reasons mentioned earlier,
the transmission matrix with odd labels $^{\rm o}T(\theta)$ refers to a situation
which is unstable and,  even if it can be defined, its properties are likely
to differ from those of $^{\rm e}T(\theta)$. The expressions \eqref{summary} are
similar to, but not quite the same as, the ansatz proposed by Konik and LeClair
\cite{Konik97}.

\p
The usual bulk bound-state bootstrap operates, presumably, just with the left to right
transmission matrix and will offer some further constraints \cite{Zam79}. Suppose the pair of
particles $a,b$ can form a bound state
$c$ at the rapidities (usual conventions)
\begin{equation}\label{}
    \theta_a=\theta_c + i\bar U_{a\bar c}^{\bar b},\quad \theta_b=\theta_c -
    i \bar U_{b\bar c}^{\bar a},\quad \bar U = \pi - U,
\end{equation}
then the heuristics will demand that the transmission matrices are consistent with this.
Thus,\begin{equation}\label{Tbootstrap}
    c_{ab}^f\, ^{\rm e}T_{f\alpha}^{c\beta}(\theta_c)= \, ^{\rm e}T_{b\alpha}^{d\gamma}(\theta_b)
    \, ^{\rm e}T_{a\gamma}^{e\beta}(\theta_a)\, c_{de}^c,
\end{equation}
where the coupling constants are denoted $c_{ab}^f$, and repeated indices are summed
(as usual). Typically, this can be used to generate the transmission matrices for
breathers and indeed that calculation will be performed later. However, it can also be
used to discuss the
`annihilation pole' at which
a particle and anti-particle virtually annihilate to the vacuum. This happens when
$b=\bar a$ and $\theta_a=\theta+i\pi/2,\ \theta_{\bar a}=\theta-i\pi/2$, and requires,
\begin{equation}\label{annihilation}
    \delta_\alpha^\beta = \sum_c\, ^{\rm e}T_{\bar a \alpha}^{c\gamma}\left(\theta-
    \frac{i\pi}{2}\right) \, ^{\rm e}T_{a\gamma}^{\bar c \beta}\left(\theta+
    \frac{i\pi}{2}\right).
\end{equation}
Since the only possibility of annihilating to the vacuum in the present case
must involve a soliton and an anti-soliton, the couplings $c_{a\bar a}^0,\ c_{c\bar c}^0$
cancel. To see the consequences of \eqref{annihilation} it is convenient to shift
$\theta$ by $i\pi/2$ and note $x(\theta+i\pi)=qx(\theta)$. Then, with the components of
the transmission matrix given by the expressions \eqref{ABCD} and \eqref{summary},
the off-diagonal components of \eqref{annihilation} are identically satisfied for either
choice of $\epsilon$, while
the two diagonal entries cannot be satisfied for the choice $\epsilon=1$.
For the other choice, $\epsilon=-1$, the diagonal entries each lead to the same condition,
namely,
\begin{equation}\label{annihilationcondition}
    \left(\nu+\frac{\lambda\mu q x^2}{\sigma^2}\right)A(\theta)A(\theta+i\pi)=1.
\end{equation}
 Hence, setting
$A(\theta)=f(q,x)/\sqrt{\nu}$, one needs to solve
\begin{equation}\label{fequation}
    (1+p^2\, x^2)f(q,x)f(q,qx)=1, \qquad p^2=\lambda\mu q/\nu\sigma^2.
\end{equation}

\p At this point it is worth returning to the unitarity condition, equation \eqref{Tevenunitarity},
which, on
choosing the above solutions with $\epsilon=-1$, requires
\begin{equation}\label{Aunitarity}
    \bar \sigma = 1/\sigma, \quad \lambda = \bar \mu\nu\sigma^2/q, \quad p^2=\bar\mu\mu,
    \quad \bar\nu\nu=1,\quad
     \bar A(\theta) A(\theta) (1+\bar\mu\mu x^2)=1.
\end{equation}
This follows immediately and the details will be omitted.
Thus, in terms of $f(q,x)$, the two equations which must be satisfied simultaneously are:
\begin{equation}\label{fequations}
    f(q,x)f(q,qx)(1+ p^2 x^2)=1= \bar f(q,x)f(q,x)(1+p^2 x^2), \quad px\equiv e^{\gamma(\theta -\eta)}
    \equiv e^{\gamma\tilde\theta}.
\end{equation}
Writing $p$ in terms of $\eta$ is deliberate and anticipates the eventual identification
of this parameter with the defect parameter described in the introduction.
As an aside, for $\epsilon=1$ the off-diagonal terms in \eqref{Tevenunitarity} cannot be satisfied,
and therefore this possibility is ruled out by the unitarity condition alone.
Henceforth, for both of the above reasons, it will be assumed $\epsilon =-1$.

\p
Clearly, \eqref{fequations} implies
\begin{equation}\label{}
    \bar f(q,x)=f(q,qx),
\end{equation}
and in essence these are the same equations as those solved by Konik and LeClair. It
is convenient to write
\begin{equation}\label{ftog}
    f(q,x)=\frac{e^{-\gamma\tilde\theta/2}\, e^{i\pi\gamma/4}}{\sqrt{2\pi}}\, g(q,x)
\end{equation}
so that
\begin{equation}\label{}
    g(q,x)\bar g(q,x)=\frac{\pi}{\cosh \gamma\tilde\theta},\quad \bar g(q,x)=g(q,qx).
\end{equation}
A solution to the latter pair of equations is given by
\begin{equation}\label{newc}
    g(q,x)=\Gamma(1/2- i\gamma\tilde\theta/\pi)
   \, \prod_{k=1}^\infty\, \frac{\Gamma(1/2 +2k\gamma-
    i\gamma\tilde\theta/\pi)\Gamma(1/2 +(2k-1)\gamma+
    i\gamma\tilde\theta/\pi)}{\Gamma(1/2 +2k\gamma+
    i\gamma\tilde\theta/\pi)\Gamma(1/2 +(2k-1)\gamma-
    i\gamma\tilde\theta/\pi)}.
\end{equation}
On the other hand, the
`minimal' solution given by Konik and LeClair can be written in the following way:
\begin{equation}\label{KLf}
    f(q,x)=\frac{e^{i\pi(1+\gamma)/4}}{(1+ipx)}\,\frac{r(x)}{\bar r(x)},
\end{equation}
where\begin{equation}\label{KLc}
    r(x)=\prod_{k=0}^\infty\frac{\Gamma(k\gamma+1/4 - i\gamma\tilde\theta/2\pi)\,
    \Gamma((k+1)\gamma+3/4 - i\gamma\tilde\theta/2\pi)}
    {\Gamma((k+1/2)\gamma+1/4 - i\gamma\tilde\theta/2\pi)
   \, \Gamma((k+1/2)\gamma+3/4 - i\gamma\tilde\theta/2\pi)},
\end{equation}
and $\tilde\theta=\theta-\eta$.
Therefore,
\begin{equation}\label{A}
    A(\theta)=\frac{1}{\sqrt{\nu}}\frac{e^{i\pi(1+\gamma)/4}}{(1+ipx)}\,\frac{r(x)}{\bar r(x)}.
\end{equation}

\p If desired, this may be written in a more compact form obtained by
making repeated use of Legendre's formula
\begin{equation}\label{Legendreformula}
    \Gamma(z)=\frac{2^{z-1/2}}{\sqrt{2\pi}}\,\Gamma\left(\frac{z}{2}\right)\,
    \Gamma\left(\frac{z}{2}+\frac{1}{2}\right).
\end{equation}
Thus, for example,
\begin{equation}\label{newc}
    \frac{r(x)}{\bar r(x)}=\frac{\Gamma(1/4- i\gamma\tilde\theta/2\pi)}
    {\Gamma(1/4+ i\gamma\tilde\theta/2\pi)}\, \prod_{k=1}^\infty\, \frac{\Gamma(1/2 +2k\gamma-
    i\gamma\tilde\theta/\pi)\Gamma(1/2 +(2k-1)\gamma+
    i\gamma\tilde\theta/\pi)}{\Gamma(1/2 +2k\gamma+
    i\gamma\tilde\theta/\pi)\Gamma(1/2 +(2k-1)\gamma-
    i\gamma\tilde\theta/\pi)}.
\end{equation}

\p It is clear from what has been said so far that the above arguments could be repeated
for a transmission matrix with odd defect labels. However, to do so might be misleading
and in particular the unitarity relations are inappropriate since it is expected,
as already mentioned, that the odd labelled defect is unstable.

\p Since the parameter $p=|\mu|=e^{-\gamma\eta}$ will turn out to be important it is worth
resummarising the
solution for the components of $^{\rm e}T$:
\begin{eqnarray}\label{Tsummary}
    A_\alpha^\beta =Q^{\, \alpha}A(\theta)\,\delta_\alpha^\beta,\quad &&D_\alpha^\beta=
    \nu Q^{-\alpha}A(\theta)\, \delta_\alpha^\beta,\quad\\
 \nn    B^\beta_\alpha=\tilde\lambda \sigma^\alpha p\, x_\theta A(\theta)\,\delta_\alpha^{\beta-2},
 \quad &&C^\beta_\alpha=\tilde\mu
 \sigma^{-\alpha}p\, x_\theta A(\theta) \,\delta_\alpha^{\beta+2},
\end{eqnarray}
where $\mu=p\tilde\mu$ and $\tilde\lambda=\nu\sigma^2/q\tilde\mu$. The phases $\nu,\tilde\mu,\sigma$
are all independent of $\theta$ and, if they are also independent of $\eta$, the transmission
matrix turns out to be a function only of the difference $\theta -\eta$.

\p
One further remark before proceeding. It is possible to make a change of basis, redefining
the transmission matrix via a diagonal unitary transformation without upsetting the
assumptions made at the beginning of the section. This must be a transformation on the defect labels
alone, and it may be written in the form
\begin{equation}\label{unitary change}
    V_\alpha^\beta = v_\alpha\delta_\alpha^\beta, \quad |v_\alpha|=1,
\end{equation}
for which $A$ and $D$ are unchanged, but $B$ and $C$ alter according to,
\begin{equation}\label{BCchanges}
    b_\alpha \rightarrow v_\alpha b_\alpha \bar v_{\alpha+2},\quad c_\alpha\rightarrow v_\alpha
    c_\alpha \bar v_{\alpha-2}.
\end{equation}
Such a transformation can be designed to remove the phases $\tilde\lambda\sigma^\alpha$ from
$b_\alpha$
by selecting
\begin{equation}\label{}
    \nn v_\alpha = \sigma^{\alpha(\alpha-2)/4 }\, \tilde\lambda^{\alpha/2},
\end{equation}
and then
\begin{equation}\label{}
    \nn c_\alpha \rightarrow \tilde\mu\tilde\lambda\sigma^{-2}= \nu/q.
\end{equation}
A further transformation of the same type can then be made to ensure the coefficients
in the off-diagonal terms are the same.
In other words, the dependence on $\sigma$ and the two phases $\tilde\mu,\ \tilde\lambda$
can be removed entirely to leave the symmetrical expression
\begin{eqnarray}\label{Tfinalsummary}
  \left(%
\begin{array}{cc}
  A_\alpha^\beta &  B^\beta_\alpha \\
   C^\beta_\alpha & D_\alpha^\beta\\
\end{array}%
\right)=f(q,x)\left(%
\begin{array}{cc}
  \nu^{-1/2}Q^{\, \alpha}\,\delta_\alpha^\beta & q^{-1/2}
  \,e^{\gamma(\theta-\eta)}\,\delta_\alpha^{\beta-2}\\
  q^{-1/2}\, e^{\gamma(\theta-\eta)}
   \,\delta_\alpha^{\beta+2} & \nu^{1/2} Q^{-\alpha}\, \delta_\alpha^\beta\\
\end{array}%
\right)
\end{eqnarray}

\p
The result \eqref{Tfinalsummary} has been arrived at via a collection of arguments based
on the triangle relation \eqref{STT} and general principles. Before proceeding to investigate
some of its consequences, it is worth noting some general features and comparing the result with
what one might have expected
on the basis of the Lagrangian description. If $\eta>0$ and $\theta>\eta$,
the character-changing elements
dominate while for $\theta<\eta$ they are suppressed. On the other hand, for $\eta<0$, the
character-changing processes are always dominant. All these facts are compatible with the classical
transmission properties described in section (2). Also, it is striking that the matrix elements
representing a soliton converting to an anti-soliton or vice-versa are the same in this basis,
while elements representing the scattering of solitons or anti-solitons with no change of character
are different. This aspect follows from an argument based on the Lagrangian and makes use of the
particular type of defect described there.

\p
Suppose the transmission matrix is expressed formally as a functional integral over the fields
$u$ and $v$,  weighted by the classical action including the defect contribution \eqref{sGdefect};
and further suppose that a defect on its own is labelled by the vacuum configurations of the fields
to either side of it. Thus the label $(a,b)$ is ascribed to the defect when the fields have
the constant values $u=2a\pi/\beta,\ v=2b\pi/\beta$. Provided the labels range over even integers
the defect potential and the bulk terms have the values they would have for $(a,b)=(0,0)$,
corresponding to a stable defect ($\sigma>0)$. The topological charge of the defect is $b-a$.
The term in the Lagrangian linear in the
time derivatives has a consequence that may be explored as follows. Field configurations
$u,v$ evolving in the presence of an initial defect with labels $(a,b)$ may be compared with
configurations
evolving with an initial defect labelled $(0,0)$ by translating the fields via
$u\rightarrow
u-2a\pi/\beta,\ v\rightarrow v-2b\pi/\beta$. The bulk and defect potential parts
do not change under this but, due to the terms linear in time derivatives, the action changes
by a term
\begin{equation}\label{actionchange}
   \frac{\pi}{\beta} \int_{-\infty}^\infty\, dt \left(av_t-bu_t\right)_{x=0}=
   \frac{\pi}{\beta}(a\delta v -b \delta u),
\end{equation}
where $\delta u,\ \delta v$ are the nett changes over time of the field configurations
evaluated at the location of the defect. Consequently, the functional integral representing
the transmission factor relative to an initial defect $(a,b)$ will differ from the
transmission relative to $(0,0)$ by a constant factor given by
\begin{equation}\label{factor}
   T(a,b)= \exp\left(\frac{i\pi}{\beta}(a\delta v -b\delta u)\right)\, T(0,0).
\end{equation}
For example, a soliton travelling in a positive sense along the $x-$axis will produce the shifts
$(a,b)\rightarrow (a-1,b-1),\ \delta u = -2\pi/\beta = \delta v $, if its character does not change,
 thus acquiring a
factor
\begin{equation}\label{}
e^{2i\pi^2(b-a)/\beta^2}\equiv Q^{\frac{(b-a)}{2}}.
\end{equation}
On the other hand, it requires the shifts
$(a,b)\rightarrow (a-1,b+1),\ \delta u = -2\pi/\beta = -\delta v $ and acquires a factor
\begin{equation}\label{}
e^{2i\pi^2(b+a)/\beta^2}\equiv Q^{\frac{(b+a)}{2}},
\end{equation}
if its character changes. The corresponding factors for an anti-soliton are, respectively,
\begin{equation}\label{}
Q^{\frac{-(b-a)}{2}} \quad\hbox{and}\quad Q^{-\frac{(b+a)}{2}}.
\end{equation}
Previously, the defect was labelled by $\alpha=b-a$, its topological charge,
but the above remarks suggest there might also be a dependence on the quantity
$p=a+b$. Thus, in the notation introduced at the beginning of this section,
in \eqref{ABCD}, for the components of $^{\rm e}T$,
it is tempting to write
\begin{eqnarray}\label{newABCD}
\nn    A_{\alpha\,p}^{\beta\, q}=Q^{\alpha/2}\, \hat a_0\,\delta_\alpha^\beta\,\delta_p^{q+2}\, ,\ &&
    B_{\alpha\,p}^{\beta\, q}=Q^{p/2}\,\hat b_0\,\delta_\alpha^{\beta-2}\,\delta_p^{q}\, ,\\
    C_{\alpha\,p}^{\beta\, q}=Q^{-p/2}\,\hat c_0\,\delta_\alpha^{\beta+2}\,\delta_p^{q}\, ,&&
    D_{\alpha\,p}^{\beta\, q}=Q^{-\alpha/2}\,\hat d_0\,\delta_\alpha^\beta\,\delta_p^{q-2}\, .
\end{eqnarray}
The expressions \eqref{newABCD} appear to have a different dependence on $Q$ to that reported in
\eqref{Tfinalsummary} and include coupling-dependent terms sensitive to $p=a+b$. However,
 the dependence on $Q$ may be adjusted using the diagonal unitary transformation
 \begin{equation}\label{}
    V_{\alpha p}^{\beta q}=Q^{p\alpha/4}\, \delta_\alpha^\beta\, \delta_p^q,
\end{equation}
to obtain instead
\begin{eqnarray}\label{newnewABCD}
\nn    A_{\alpha\,p}^{\beta\, q}=Q^\alpha\, \hat a_0\,\delta_\alpha^\beta\,\delta_p^{q+2}\, ,&&
    B_{\alpha\,p}^{\beta\, q}=\hat b_0\,\delta_\alpha^{\beta-2}\,\delta_p^{q}\, ,\\
    C_{\alpha\,p}^{\beta\, q}=\hat c_0\,\delta_\alpha^{\beta+2}\,\delta_p^{q}\, ,\ \ \ \ &&
    D_{\alpha\,p}^{\beta\, q}=Q^{-\alpha}\,\hat d_0\,\delta_\alpha^\beta\,\delta_p^{q-2}\, .
\end{eqnarray}
Using this as a starting point and applying the same arguments as before leads to a similar
expression to \eqref{Tfinalsummary}, the only changes being the extra Kronecker $\delta$'s needed to
record the changes in $a+b$ as a soliton or anti-soliton passes the defect. In fact,
a unitary basis can be found which removes from this representation even the dependence on $\nu$,
the undetermined parameter in \eqref{Tfinalsummary}.
On the other hand, although the transmission matrix has not been defined explicitly
by a functional integral, the arguments given above confirm both the importance of the
term linear in time derivatives and the precise dependence on the coupling that has
been coded into $Q$. As a consequence, if it ever becomes possible to give an
independent derivation of the
transmission matrix then  the triangle relations \eqref{STT} could be used to provide an
alternative derivation of the soliton S-matrix.

\section{Unstable bound states as poles in $^{\rm e}T$}

\jumpup
Returning to an earlier line of thought, consider the poles of
$^{\rm e}T(\theta)$.  At first sight it seems from \eqref{KLf} that each
component of $^{\rm e}T(\theta)$ has a pole at
$px=i$, or in terms of rapidity at $\theta = \eta +i\pi/2\gamma$ (or, equivalently,
$i\gamma\tilde\theta/2\pi = -1/4$). However, this is an
illusion since $\bar r(x)$ also has a  pole at the same location which cancels it out.
On the other hand, it is evident from the expressions \eqref{newc} and \eqref{ftog}
that there is a pole at $\theta = \eta -i\pi/2\gamma$ (that is, $i\gamma\tilde\theta/\pi = 1/2$),
and this pole does not have a compensating zero. If this pole is taken to
be significant then it corresponds to a defect `bound' state with an energy
jump given by
\begin{equation}\label{energyjump}
    \Delta E=E_b-E_0=m_s\cosh\left(\eta-\frac{i\pi}{2\gamma}\right)=m_s\cosh \eta\,
    \cos\left(\frac{\pi}
    {2\gamma}\right)-im_s\sinh \eta\, \sin\left(\frac{\pi}{2\gamma}\right),
\end{equation}
where $m_s$ is the mass of the soliton or anti-soliton.
Equation \eqref{energyjump} is  interesting because in view of \eqref{gamma},
the limit $\beta
\rightarrow 0$ implies $1/\gamma \rightarrow 0$ and the real part of $\Delta E$ becomes the
classical energy of a soliton with
rapidity $\eta$, and the imaginary part of $\Delta E$ vanishes. This suggests this pole
corresponds to the
absorption and emission of a soliton (or anti-soliton) and the imaginary part of
the pole location governs
the width of the `bound' state (in other words its decay,  corresponding to emission).
The sign of the imaginary part is reasonable since one would expect the
wave-function of the bound state to evolve according to
\begin{equation}\label{}
   \psi_b(t)= e^{-iE_b\, t}\psi_b(0).
\end{equation}
Thus, $\psi_b(t)$ is certainly decaying with time provided the imaginary part of
$\Delta E$ is negative.
In the classical limit the width goes to zero so the
only process is absorption. From this point of view, it is fortunate the other pole
turned out to be a phantom.

\p
Another interesting point concerns the picture if momentum is incorporated. Since a particle
generally passes through a defect, and in the classical picture can exchange both
energy and momentum with it, it becomes clear from the expressions
for the energy and momentum that the energy-momentum of the excited
quantum defect is a complex rotation of the energy momentum of a bulk soliton with
rapidity $u$. Thus,
\begin{equation}\label{}
    \left(%
\begin{array}{c}
  \Delta E\\
  \Delta P\\
\end{array}%
\right)
=\left(%
\begin{array}{cc}
  \cos\chi & -i\sin\chi\\
   -i\sin\chi& \cos\chi\\
\end{array}%
\right)\left(%
\begin{array}{c}
  m_s\cosh \eta \\
  m_s\sinh \eta\\
\end{array}%
\right),\quad \chi = \frac{\pi}{2\gamma}.
\end{equation}
Clearly,
\begin{equation}\label{}
    (\Delta E)^2-(\Delta P)^2=m_s^2,
\end{equation}
and hence, despite having a complex energy and momentum suited to representing an unstable
state, the excited state remains on the soliton mass shell. This seems to
fit exceedingly well with the classical picture of a defect (of odd topological charge
and therefore additional energy) being a `hidden' soliton.

\p Another interesting limit corresponds to $\beta^2=4\pi$ ($\gamma=1$). In that limit,
the solitons become free \cite{Zam79},  and  the real parts of $\Delta E$ and
$\Delta P$ vanish.
Nevertheless, the bound state remains on the soliton mass shell. In that limit, the r\^oles
of energy and momentum have completely interchanged and, in that sense, one might
wonder about an analogy with a black hole described by the Schwarzschild metric
for which `timelike' and `spacelike' interchange roles at the horizon.
In this limit, $Q^2=1$ and the
transmission factor \eqref{Tfinalsummary} simplifies considerably to
\begin{equation}\label{freefermionT}
    ^{\rm e}T_{a\, \alpha}^{b\, \beta}(\theta)= \frac{1}{(1-ie^{\tilde\theta})}
    \left(%
\begin{array}{cc}
 \nu^{-1/2} \delta_\alpha^\beta & -i\,e^{\tilde\theta}\,\delta_\alpha^{\beta-2}\\
  -i\, e^{\tilde\theta}\,\delta_\alpha^{\beta+2} &\nu^{1/2}\, \delta_\alpha^\beta \\
\end{array}%
\right).
\end{equation}

\p
Near the bound state pole, the transmission matrix should have the form
\begin{equation}\label{pole}
    ^{\rm e}T_{a\alpha}^{b\beta}(\theta)\sim \frac{it_{a\alpha}^{b\beta}}
    {\left(\theta-\eta+\frac{i\pi}{2\gamma}\right)}\equiv
    \frac{i\,c_{a\alpha}^\delta\,\tilde{c}_\delta^{\,b\beta}}
    {\left(\theta-\eta+\frac{i\pi}{2\gamma}\right)}\, ,
\end{equation}
where $c_{a\alpha}^\delta$ and $\tilde{c}_\delta^{\,b\beta}$ are a pair of
coupling matrices factorizing $t_{a\alpha}^{b\beta}$. Note that while $\alpha$ and $\beta$
are even, the internal sum implied by the repeated $\delta$ is over the odd integers.
 However, there is clearly considerable freedom in choosing these couplings.
An explicit calculation of the pole residue reveals  - in the notation of
\eqref{Tfinalsummary},
\begin{equation}\label{poleresidue}
    ^{\rm e}T_{a\alpha}^{b\beta}(\theta)\sim \frac{1}{\gamma}
   \, \frac{e^{i\pi(1+\gamma)/4}}{(\theta-\eta+\frac{i\pi}{2\gamma})}\,
   \left(%
\begin{array}{cc}
 i\nu^{-1/2}Q^\alpha\, \delta_\alpha^\beta&   q^{-1/2}\,\delta_\alpha^{\beta-2} \\
   q^{-1/2}\, \delta_\alpha^{\beta+2} &  i\nu^{1/2} Q^{-\alpha}\,\delta_\alpha^\beta
  \\
\end{array}%
\right)\, .
\end{equation}
Some care needs to be taken when manipulating the infinite products of gamma functions
but the result \eqref{poleresidue} can be checked with the special case $\gamma=1$.

\p
The intermediate state, of complex energy $E_b$, if this is
a correct interpretation, should differ from the initial state of energy $E_0$ by a single
unit of topological charge. For this reason, there will be a  bootstrap condition
linking the even and odd transmission matrices. In detail, it ought to
read \begin{equation}\label{newbootstrap}
    c_{b\alpha}^\gamma \, ^{\rm o}T_{a\gamma}^{c\delta}=S_{ab}^{pq}
    \left(\theta-\eta+\frac{i\pi}{2\gamma}\right)
    \, ^{\rm e}T_{q\alpha}^{c\beta}\, c_{p\beta}^\delta,
\end{equation}
where, for example, $\alpha,\,\beta$ are both even, and $\gamma,\, \delta$
are both odd. One can argue to this assignment by considering the classical
energy of the defect poised to emit a soliton: the odd-charged defect has higher energy
and hence it is to be expected that the above pole really does refer to a pole in the
transmission matrix $^{\rm e}T$ for a particle passing through an even-charged defect. Then it
is clear $^{\rm o}T$ refers to a particle scattering with an odd-charge defect. It is
worth pointing out that the bootstrap
relation and the Yang-Baxter equation satisfied by the $S$-matrix, along with
\eqref{STT}, will guarantee that  $^{\rm o}T$ satisfies a corresponding set of equations
to \eqref{STT}, and therefore the general solution will be of a similar form to \eqref{summary}.
The corresponding quantities will be denoted by $\hat A(\theta),\ \hat\nu,$ etc. However,
although the equation corresponding to \eqref{annihilationcondition} is expected to be satisfied,
it is not expected there will be an analogue of the unitarity condition \eqref{Tevenunitarity}.

\p In order to facilitate the checking of \eqref{newbootstrap} it will be necessary to
calculate the couplings $c_{a\alpha}^\beta$ and to evaluate the sine-Gordon S-matrix at
a special point. The simplest example of \eqref{newbootstrap} concerns a soliton scattering
with a soliton-defect bound state since then one must have,
\begin{equation}\label{solitonsolitoncase}
    c_{+\alpha}^\gamma \, ^{\rm o}T_{+\gamma}^{+\delta}(\theta)= S_{++}^{++}
    \left(\theta - \eta +\frac{i\pi}{2\gamma}\right)\,
    ^{\rm e}T_{+\alpha}^{+\beta}c_{+\beta}^\delta,
\end{equation}
which in terms of the specific entries of the transmission matrices translates into
\begin{equation}\label{}
    c_{+\alpha}^\gamma \,  \hat a_\gamma =S_{++}^{++}
    \left(\theta - \eta +\frac{i\pi}{2\gamma}\right) \, a_\alpha c_{+\alpha}^\gamma.
\end{equation}
However, on the left hand side it is clear $\gamma=\alpha+1$ and so the couplings cancel
(assuming they are invertible matrices) and
\begin{equation}\label{otheraequation}
    Q^{\alpha+1}\hat A(\theta)=S_{++}^{++}
    \left(\theta - \eta +\frac{i\pi}{2\gamma}\right) Q^\alpha A(\theta).
\end{equation}
 Since, the
right hand side is in principle known, the left hand side can be calculated. If this idea is to
be consistent there must be some very interesting identities involving the S-matrix elements
evaluated at a rapidity shifted by the position of the bound state pole.  For example, it needs
to be checked that this relation is consistent with what has been assumed already concerning
the transmission matrix. The first useful identity satisfied by the S-matrix is
\begin{equation}\label{aidentity}
    S_{++}^{++}(\theta-\eta+i\psi)\, S_{++}^{++}(\theta-\eta+i\psi + i\pi)=-q\,
    \frac{1-e^{2\gamma(i\psi- \eta)}\, x^2}
    {1- e^{2\gamma(i\psi-\eta)}\,q^2 x^2},
\end{equation}
where $\eta,\psi $ are real. On the other hand, \eqref{otheraequation} implies
\begin{equation}\label{}
    Q^2 \, \hat A(\theta)\, \hat A(\theta+i\pi)=S_{++}^{++}(\theta-\eta+i\psi)
    S_{++}^{++}(\theta-\eta+i\psi + i\pi)
 A(\theta)A(\theta+i\pi),
\end{equation}
and hence, using \eqref{annihilationcondition}, one finds
\begin{equation}\label{aoddconstraint}
    \frac{1}{\, \hat \nu(1+\hat p^2x^2)}=\left(\frac{1-e^{2\gamma(i\psi- \eta)}\, x^2}
    {1- e^{2\gamma(i\psi-\eta)}\,q^2 x^2}\right)\,  \frac{1}{ \nu(1+ p^2x^2)}.
\end{equation}
Since $x=e^{\gamma\theta}$ is free, comparing powers of $x$ leads to the conclusions:
\begin{equation}\label{hatp}
    \hat\nu =\nu, \quad \hat p^2 = q^2 p^2,\quad \psi=\frac{\pi}{2\gamma}
    \pm \frac{k\pi}{\gamma},\ k\ {\rm integer}.
\end{equation}
An explicit computation of $\hat A(\theta)$ using \eqref{otheraequation} yields,
\begin{equation}\label{Ahat}
    \hat A(\theta)=\frac{1}{\sqrt{\nu}}\, \frac{e^{-i\pi(1+\gamma)/4}}{(1+ipx)}\,
    \frac{\cos\left(\pi/4\gamma-i(\theta-\eta)/2\right)}
    {\sin\left(\pi/4\gamma-i(\theta-\eta)/2\right)}\, \frac{s(x)}{\bar{s}(x)},
\end{equation}
where $s(x)$ is quite similar to $r(x)$ defined in \eqref{KLc}
\begin{equation}\label{KLs}
    s(x)=\prod_{k=0}^\infty\frac{\Gamma(k\gamma+3/4 + i\gamma\tilde\theta/2\pi)\,
    \Gamma((k+1)\gamma+1/4 + i\gamma\tilde\theta/2\pi)}
    {\Gamma((k+1/2)\gamma+1/4 + i\gamma\tilde\theta/2\pi)
   \, \Gamma((k+1/2)\gamma+3/4 + i\gamma\tilde\theta/2\pi)}.
\end{equation}
Using \eqref{Ahat} it is not difficult to check directly that the condition
\eqref{annihilationcondition} holds but with $p$ replaced by $\hat p$, namely
\begin{equation}\label{}
    \nn \hat A(\theta)\hat A(\theta+i\pi)=\frac{1}{\nu\, (1+\hat p^2x^2)}.
\end{equation}
Note, no further information is obtained from considering an anti-soliton scattering with an
anti-soliton-defect bound state since that would require
\begin{equation}\label{}
    c_{-\alpha}^\gamma \, \hat d_\gamma =S_{--}^{--}
    \left(\theta - \eta +\frac{i\pi}{2\gamma}\right)\, d_\alpha c_{-\alpha}^\gamma.
\end{equation}
On the left hand side $\gamma = \alpha -1$ and therefore, recalling the properties
of the S-matrix and that $d_\alpha=\nu Q^{-\alpha}
A(\theta)$, one finds
\begin{equation}\label{}
    Q\hat \nu \hat A(\theta)=S_{++}^{++}\left(\theta - \eta +\frac{i\pi}{2\gamma}\right)
    \nu A(\theta).
\end{equation}
This is exactly the previous relation because $\hat\nu = \nu$.
Clearly, from the relations \eqref{hatp}, when there is a complex bound state pole
not both of $p$ and
$\hat p$ can be real.
In turn this confirms that the unitarity assumption must be faulty for $^{\rm o}T(\theta)$,
as  expected. On the other hand, the relationship corresponding to
\eqref{annihilationcondition} survives. In fact, the incompatibility of
\eqref{otheraequation} with unitarity can be checked directly by multiplying each side
of \eqref{otheraequation} by its complex conjugate and comparing with \eqref{Aunitarity}.
The fact the S-matrix
is evaluated at $\theta - \eta +i\pi/2\gamma$ provides an immediate contradiction since
$$\bar S_{++}^{++}(\theta - \eta +i\pi/2\gamma) S_{++}^{++}(\theta - \eta +i\pi/2\gamma)\ne 1.$$
To obtain further information from \eqref{newbootstrap} would require more complete knowledge
of the coupling matrices.

\section{Transmission factors for the breathers}
\jumpup
In this section the precise form of the transmission factor for the lightest breather will be
calculated and it turns out to be surprising. Knowledge of this particular transmission factor
should provide a way to begin checking perturbatively the
properties of the defect: since the lightest breather corresponds to the
quantum particle described by the fundamental bulk scalar field, a development of
standard methods can be used to examine low orders of perturbation theory.

\p
The breather bound state poles of the soliton-soliton scattering matrix \eqref{Smatrix}
occur at \cite{Zam79}
\begin{equation}\label{breatherpoles}
    \Theta = i\left(\pi-\frac{n\pi}{\gamma}\right),\quad n=1,2,3,....,n_{\rm max},
\end{equation}
where $n_{\rm max}$ depends upon the coupling $\beta$ and is defined as the largest integer
for which $n/\gamma<1$. The corresponding breather masses are given by
\begin{equation}\label{}
    m_n=2m_s \sin\left(\frac{n\pi}{2\gamma}\right).
\end{equation}
The existence of the breather poles allows a calculation of breather transmission factors
$^nT(\theta)$ via a bootstrap relation as follows \cite{Konik97}
\begin{equation}\label{breatherbootstrap}
    c_{a\bar a}^{n}\, ^nT(\theta)\, \delta_\alpha^\beta =\sum_b\,
    ^{\rm e}T_{\bar a \alpha}^{\bar b \gamma}
    (\theta_{\bar a})\, ^{\rm e}T_{ a \gamma}^{ b \beta}(\theta_{a})\, c_{b\bar b}^{n},
\end{equation}
where
\begin{equation}\label{}
 \theta_a=\theta+i\left(\frac{\pi}{2}-\frac{n\pi}{2\gamma}\right), \quad  \theta_{\bar a}=
 \theta -i\left(\frac{\pi}{2}-\frac{n\pi}{2\gamma}\right),
\end{equation}
and the label $b$ is summed over the two possibilities $b=\pm$. It is necessary to be slightly
careful with the couplings in \eqref{breatherbootstrap} because
$$c_{+-}^{n}=(-)^nc_{-+}^{n},$$
especially since the main interest here lies with the first breather $(n=1)$. Taking that into
account and using the explicit expressions for the soliton transmission matrix elements
given in \eqref{summary}
the breather transmission matrix for the lightest breather is given by
\begin{equation}\label{Tbreather}
    ^1 T(\theta) \delta_\alpha^\beta = \left( D(\theta_{\bar a})A(\theta_{a})-
    C(\theta_{\bar a})B(\theta_{a})\right)_\alpha^\beta = \left(1-\frac{p^2x^2}{q}\right)
    \nu A(\theta_{\bar a})A(\theta_a)\delta_\alpha^\beta.
\end{equation}
It is straightforward to use the expression for $A(\theta)$ provided by \eqref{KLf}
to calculate $^1T(\theta)$, the only additional formula needed being the infinite
product identity for the sine function
\begin{equation}\label{sineproduct}
    \frac{\sin \pi z}{\pi z}=\prod_{k=1}^\infty \left(1-\frac{z^2}{k^2}\right).
\end{equation}
The result is surprising since all dependence on the bulk coupling cancels out to
leave:
\begin{equation}\label{Tbreatherexplicit}
    ^1T(\theta)= i\, \frac{\sin\left(\frac{\pi}{4}+\frac{i(\theta-\eta)}{2}\right)}
{\sin\left(\frac{\pi}{4}-\frac{i(\theta-\eta)}{2}\right)}\equiv -i\,
\frac{\sinh\left(\frac{\theta-\eta}{2}-\frac{i\pi}{4}\right)}
    {\sinh\left(\frac{\theta-\eta}{2}+\frac{i\pi}{4}\right)}.
\end{equation}
Moreover,  this expression is  identical to the transmission factor
in the classical (linear) limit given in section (2) by \eqref{classicalT}.
Given the delicate cancellations this seems a noteworthy result.

\p Given that the sinh-Gordon model is the imaginary coupling version of the
sine-Gordon model, the result \eqref{Tbreatherexplicit} also suggests that the scalar
particle of the sinh-Gordon model has a transmission factor independent of the
bulk coupling.

\p Although the result \eqref{Tbreatherexplicit} is surprising, it might be altered
if the `minimal'
version of $^{\rm e}T(\theta)$ were to be modified by the addition of CDD factors.
The purpose of a perturbative analysis would be to establish the validity, or
otherwise, of \eqref{Tbreatherexplicit}, and hence to determine whether or not
 CDD factors might be necessary. Resolving this issue will be postponed to
 a future report.

 \p The transmission factors for the other breathers may be obtained iteratively
 using the bootstrap. For example, denoting the scattering matrix of the
 lightest breather by $S_{11}(\Theta)$,
 \begin{equation}\label{}
    S_{11}(\Theta)= -\frac{\sinh\left(\frac{\Theta}{2}+\frac{i\pi}{2\gamma}\right)
    \sinh\left(\frac{\Theta}{2}-\frac{i\pi}{2}\left(1+\frac{1}{\gamma}\right)\right)}{
    \sinh\left(\frac{\Theta}{2}-\frac{i\pi}{2\gamma}\right)
    \sinh\left(\frac{\Theta}{2}+\frac{i\pi}{2}\left(1+\frac{1}{\gamma}\right)\right)}=
    -\,(2/\gamma)_{\Theta}(2-2/\gamma)_{\Theta},
\end{equation}
where the last expression makes use of the bracket notation
\begin{equation}\label{}
    (z)_{\theta}=\frac{\sinh\left(\frac{\theta}{2}+\frac{i\pi z}{4}\right)}
    {\sinh\left(\frac{\theta}{2}-\frac{i\pi z}{4}\right)},
\end{equation}
there is a pole corresponding to the next breather at $\Theta=i\pi/\gamma$. Using this
the transmission factor of the second breather will be given by
\begin{equation}\label{secondbreather}
    ^{\rm 2}T(\theta)=\, ^{\rm 1}T\left(\tilde\theta +\frac{i\pi}{2\gamma}\right)
   \, ^{\rm 1}T\left(\tilde\theta -\frac{i\pi}{2\gamma}\right)=
   -\,\frac{1}{(1+ 1/\gamma)_{\tilde\theta}(1-1/\gamma)_{\tilde\theta}}\, ,
\end{equation}
where $\tilde\theta=\theta-\eta$. The result is obviously dependent on the bulk coupling.
Unlike the expression \eqref{Tbreatherexplicit}
the transmission factor for the second breather contains two complex poles possibly indicating
the existence of an excited state of the defect with no change in its topological charge.
However, using the same argument as before, any defect-breather bound state would remain
on the mass shell of the breather and would need to be interpreted as the quantised version
of a phenomenon in the classical model. However, mysteriously, there do not appear to be any
periodic solutions associated specifically with the defect. On the other hand,
from an energy relation similar to \eqref{energyjump} it is clear that for the two possible
poles, $\tilde\theta = -i\pi(1\pm1/\gamma)/2$,
\begin{equation}\label{benergyjump}
    \Delta E_\pm=\mp\, m_2\cosh\eta\,\sin\left(\frac{\pi}{2\gamma}\right)-
    im_2\sinh\eta\,\cos\left(\frac{\pi}{2\gamma}\right),
\end{equation}
while the pole coefficients are  $\pm 2 i\cot(\pi/2\gamma)$, respectively. Thus, the pole with
a positive residue has ${\rm Re}\,\Delta E <0$, and the pole with
${\rm Re}\,\Delta E >0$ has a negative
residue. As $\beta\rightarrow 0$ (i.e. $1/\gamma\rightarrow 0$), the pair of poles coalesce to
a double pole. For $\gamma>1$, neither of these poles is on the `physical strip'
($0<{\rm Im}\,\tilde\theta<\pi$). Nevertheless,
it is an interesting question to decide the nature of the states indicated
by these poles, or what their origin might be if they do not correspond to unstable states.

\p The general case needs to be split into `odd' and `even' breathers, according to whether
$n$ in \eqref{breatherpoles} is odd or even. In summary, repeatedly using
\eqref{breatherbootstrap}
one finds,
\begin{equation}\label{oddbreathers}
    ^{\rm 2s+1}T(\theta)=(-)^{s+1}\frac{i}{(1)_{\tilde\theta}}\, \prod_{l=0}^{s-1}\,
    \frac{1}{(1+2(s-l)/\gamma)_{\tilde\theta}\, (1-2(s-l)/\gamma)_{\tilde\theta}}\, ,
\end{equation}
for $n$ odd, and
\begin{equation}\label{evenbreathers}
^{\rm 2s}T(\theta)=(-)^{s}\, \prod_{l=0}^{s-1}\,
    \frac{1}{(1+(2s-2l-1)/\gamma)_{\tilde\theta}\, (1-(2s-2l-1)/\gamma)_{\tilde\theta}}\, ,
\end{equation}
for $n$ even. Typically, in common with the transmission factor for the second breather,
these have complex poles suggesting the existence of additional
excited states.

\section{Scattering defects}
\jumpup
As remarked earlier in section (3) there is no reason in principle why defects
themselves will not scatter and the purpose of this section is to make some comments about
this. Throughout the section it will be assumed the defect parameter is positive, so that the
even type of defect is stable. Then, it could be expected that there would be asymptotic states
composed of any number of defects with different speeds. It might be expected the
speed of a defect will modify the defect parameter but it is necessary to redo some
classical parts of the analysis to determine if this is really correct and if so, how
the transmission matrix already determined will be modified. It was noted in \cite{bczlandau}
how the requirement of having a conserved total momentum, including a defect
contribution, was  equivalent to the requirements of integrability. In fact,
the existence of the conserved momentum implied the form of the defect boundary
conditions. This observation will be used as a shortcut in the present discussion.

\p
Starting from first principles, the action with a moving defect is taken to be
\begin{equation}\label{action}
    A=\int dt \left\{\int_{-\infty}^z dx {\cal L}(u) +
    {\cal\bf B}-m_D\sqrt{1-\dot z^2} + \int_z^\infty
    dx {\cal L}(v)\right\},
\end{equation}
where $z$ depends on time and ${\bf B}$ depends on the fields evaluated at $x=z(t)$,
and possibly on $z$ or $\dot z$ in a manner to be determined.
A velocity dependent action for the defect
is included assuming its mass to be $m_D$. On the defect, the total time
derivative of
either field (or its variation) will be given in terms of a combination such as
\begin{equation}\label{totaltderv}
   \dot u= \frac{du}{dt}=\frac{\partial u }{\partial t} + \dot z\,
   \frac{\partial u}{\partial x}
   \equiv u_t+\dot z u_x,
\end{equation}
or a similar expression for $\dot v$.

\p
Then, requiring $A$ to be stationary with respect to  variations of $u$ gives:
\begin{eqnarray}
  \frac{\partial{\cal L}}{\partial u_x}+\frac{\partial{\cal\bf B}}{\partial u}-
  \frac{d}{dt}\,\frac{\partial{\cal\bf B}}{\partial u_t}-\dot z \frac{\partial{\cal L}}
  {\partial u_t} &=& 0, \quad x=z\label{phidc}\\
  \frac{\partial{\cal \bf B}}{\partial u_x}-\dot z
  \frac{\partial{\cal \bf B}}{\partial{u_t}}&=& 0, \quad x=z\label{dphidt} \\
  \frac{\partial{\cal L}}{\partial u}-\partial_\mu\frac{\partial{\cal L}}
  {\partial u_\mu} &=& 0, \quad x<z.
\end{eqnarray}
Equation \eqref{dphidt} simply implies ${\cal\bf B}$ depends on the combination $du/dt$
while equation \eqref{phidc} gives the boundary condition for $\partial u/\partial x$:
\begin{equation}\label{}
u_x=\frac{\partial{\cal\bf B}}{\partial u}-
  \frac{d}{dt}\,\frac{\partial{\cal\bf B}}{\partial u_t}-\dot z u_t, \quad x=z;
\end{equation}
this suggests it is natural to choose as a defect contribution
\begin{equation}\label{defectterm}
    {\cal\bf B}=\frac{1}{2}\left(u\frac{dv}{dt}-v\frac{du}{dt}\right)-{\cal B}.
\end{equation}
In the latter, it may then be assumed ${\cal B}$ will have no dependence on the
derivatives of the fields (but will still depend on $z,\dot z$).

\p
Then, (and similarly for $v$),
\begin{equation}\label{conditionsx}
    u_x = \frac{1}{1-\dot z^2}\left(-\frac{\partial{\cal B}}{\partial u}+
    \frac{dv}{dt}-\dot z \frac{du}{dt}\right),\quad
    v_x=\frac{1}{1-\dot z^2}\left(\phantom{-} \frac{\partial{\cal B}}{\partial v}+
    \frac{du}{dt}-\dot z \frac{dv}{dt}\right) ,
\end{equation}
both of which reduce to the old situation when $z$ is constant.
Alternatively, one can combine these differently to obtain:
\begin{equation}\label{alternativedefectconditions}
u_x =\frac{1}{1-\dot z^2}\left(-\frac{\partial{\cal B}}{\partial u}+
\dot z \frac{\partial{\cal B}}{\partial v}
\right) + v_t, \quad
 v_x =\frac{1}{1-\dot z^2}\left(\phantom{-}
 \frac{\partial{\cal B}}{\partial v}-\dot z \frac{\partial{\cal B}}{\partial u}
\right) + u_t.
\end{equation}
For some purposes the latter is simpler.  There are similar
expressions for the partial time derivatives at the defect:
\begin{equation}\label{conditionst}
    u_t = \frac{1}{1-\dot z^2}\left(\phantom{-} \dot z\frac{\partial{\cal B}}{\partial u}+
    \frac{du}{dt}-\dot z \frac{dv}{dt}\right), \quad
    v_t =\frac{1}{1-\dot z^2}\left(-\dot z\frac{\partial{\cal B}}{\partial v}+
    \frac{dv}{dt}-\dot z \frac{du}{dt}\right) ;
\end{equation}
these are identities when $z$ is constant.

\p
Next, consider the contribution of the two bulk fields $u$ and $v$ to the total momentum,
\begin{equation}\label{}
    P=\int_{-\infty}^z dx\, u_tu_x +\int_z^\infty dx\, v_tv_x.
\end{equation}
This is not expected to be conserved. However, the time derivative of $P$
is
\begin{eqnarray}
  \dot P &=& \left[\dot z \left(u_tu_x - v_tv_x\right)
   +\frac{1}{2}\left(u_t^2 +u_x^2 - v_t^2 -v_x^2\right)
   -\left( V(u)-W(v)\right)\right]_{x=z},
\end{eqnarray}
and the various partial derivatives evaluated at the defect can all be related
to the total time derivatives using  the defect conditions for $u_x, v_x$,
and the definition of the total time derivative.
One finds that all terms quadratic in total time derivatives cancel out;
that terms linear in total derivatives reduce to
\begin{equation}\label{onederv}
    \frac{1}{1-\dot z^2}\left[\dot u\left(\dot z \frac{\partial{\cal B}}{\partial u}-
    \frac{\partial{\cal B}}{\partial v}\right)+\dot v\left(\dot z \frac{\partial{\cal B}}
    {\partial v}-\frac{\partial{\cal B}}{\partial u}\right)\right];
\end{equation}
and that terms without any derivatives at all simplify to
\begin{equation}\label{nodervs}
    \frac{1}{2(1-\dot z^2)}\left[\left(\frac{\partial{\cal B}}{\partial u}\right)^2
    -\left(\frac{\partial{\cal B}}{\partial v}\right)^2\right] - V(u) +W(v).
\end{equation}

\p
Consider first the last of these. The two potentials do not depend on $\dot z$ and, therefore,
a suitable proposal for ${\cal B}$ will be
\begin{equation}\label{}
    {\cal B} = \sqrt{1-\dot z^2}\, {\cal C},
\end{equation}
where ${\cal C}$  is chosen to ensure \eqref{nodervs} vanishes identically when $\dot z=0$.
Assuming this to be the case, \eqref{onederv} is quite close to a total time derivative.
Setting
\begin{equation}\label{}
    \frac{\partial{\cal C}}{\partial u}=\frac{\partial {\cal U}}{\partial v}, \quad
    \frac{\partial{\cal C}}{\partial v}=\frac{\partial {\cal U}}{\partial u},
\end{equation}
\eqref{onederv} becomes
\begin{equation}\label{onederva}
    \frac{1}{\sqrt{1-\dot z^2}}\left[\dot u(\dot z {\cal C}_u-{\cal U}_u)+
    \dot v(\dot z {\cal C}_v -{\cal U}_v)\right].
\end{equation}

\p
Finally, if ${\cal C}$ is chosen to have the form
\begin{equation}\label{}
    {\cal C}(\sigma)= \sigma F(u+v)+\frac{1}{\sigma}F(u-v),
\end{equation}
and $\sigma_0$ is defined by
\begin{equation}\label{zdotrelation}
    \sigma=\sigma_0 \sqrt{\frac{1+\dot z}{1-\dot z}},
\end{equation}
then \eqref{onederv} becomes a total time derivative of
$$(\dot z {\cal C}-{\cal U})/\sqrt{1-\dot z^2}.$$
Hence,
\begin{equation}\label{momentum}
    {\cal P}= P+\left[ \sigma_0
    F(u+v)-\frac{1}{\sigma_0}F(u-v)\right]_{x=z}
\end{equation}
is exactly conserved. In other words, the contribution to the total momentum of a defect
with parameter $\sigma$ is precisely
the same as it would be if the defect were at rest with parameter $\sigma_0$,
except that the fields $u$ and $v$ are evaluated at its actual location $x=z(t)$.

\p
Since time
translation invariance is unbroken energy conservation is guaranteed with
the precise expression
 given below \eqref{energy}.
Checking the energy explicitly by calculating the time derivative of the
bulk contributions $E
=E(u)+E(v)$
gives
\begin{equation}\label{}
    \dot E = \left[u_tu_x -v_tv_x +\frac{\dot z}{2} \left(u_t^2 +u_x^2
    -v_t^2 - v_x^2 +2V(u)-2W(v)\right)\right]_{x=z}.
\end{equation}
 Using the defect conditions provides a simplification of the right hand side to
\begin{equation}\label{}
    -\frac{d}{dt}\left(\sigma_0 F(u+v)+\frac{1}{\sigma_0}F(u-v)\right),
\end{equation}
implying that
\begin{equation}\label{energy}
    {\cal E}=E+\left[\sigma_0 F(u+v)+\frac{1}{\sigma_0}F(u-v)\right]_{x=z}
    \equiv E+{\cal C}(\sigma_0),
\end{equation}
is conserved. This is a little surprising since a moving defect might be expected to
have some kinetic energy associated with it, which is exchangeable with the fields.
However, as was
the case with the
momentum \eqref{momentum}, the energy \eqref{energy} is just the energy of the
defect system as if it were at rest with parameter $\sigma_0$.

\p In this notation, the potential part of the boundary term is expressed by
\begin{equation}\label{funnyB}
    {\cal B}=\sigma_0(1+\dot z)F(u+v) +\frac{1}{\sigma_0}(1-\dot z)F(u -v),
\end{equation}
where the fields are evaluated at $x=z(t)$.

\p It has already been remarked that the stationary defects exert no forces
on each other and therefore it is expected they might move freely with zero
acceleration; besides, from the above discussion, they do not exchange kinetic energy
or momentum with the fields. To check this is really consistent,
consider the equation of motion
for the defect position obtained by varying $z$. It turns out to be
\begin{equation}\label{zequation}
    \left[{\cal L}(u) - {\cal L}(v)\right]_{x=z} +
    \frac{\partial{\bf B}}{\partial z} -\frac{d}{dt}
    \frac{\partial{\bf B}}{\partial \dot z}=\frac{m_D\ddot z}{\sqrt{1-\dot z^2}}.
\end{equation}
With the choices made above
in \eqref{defectterm}, \eqref{funnyB}, and using the conditions imposed
on the fields at the defect
 by \eqref{conditionsx} and \eqref{conditionst}, the equation of motion \eqref{zequation}
reduces to $$m_D\ddot z =0.$$ Thus, it is certainly correct that the defect can move
with any chosen constant speed and explains why it was not necessary for the defect
to exchange kinetic energy with the fields on either side of the defect.
If there are several defects, the whole discussion
applies independently to each of them. However, because the
speed of each defect is arbitrary, defects will pass each other and it becomes
necessary to investigate their scattering.
If the speed of the defect is itself parameterized
using a rapidity variable $\chi$ then \eqref{zdotrelation} reads
\begin{equation}\label{parameters}
    \sigma=\exp(-\eta)=\exp(\chi-\eta_0), \quad \dot z =\tanh\chi.
\end{equation}

\p
Suppose there are two defects with parameters $\sigma_1$ and $\sigma_2$, and rapidities
$\chi_1>\chi_2$, initially separating bulk regions containing fields
$u_1,\ u_2$ and $u_2,\ u_3$,
respectively. A soliton traversing this pair from left to right will be delayed by the first
with parameter $\eta_1$ and then by the second with parameter $\eta_2$. On the other hand,
some time later the two defects will have interchanged and a similar soliton would then
first encounter the defect with parameter $\eta_2$ and then the defect with parameters
$\eta_1$.
In either case, the overall delay will be the same, and this is guaranteed by the B\"ackland
character of the defect conditions, and in fact represents a statement of Bianchi's
Theorem of Permutability.

\p
The defect conditions \eqref{alternativedefectconditions} become very simple with these
choices, namely,
\begin{equation}
u_x = - \frac{\partial{\cal C}(\sigma_0)}{\partial u}+v_t, \quad
v_x=\phantom{-}\frac{\partial{\cal C}(\sigma_0)}{\partial v}+u_t.
\end{equation}
This implies straightforwardly, at least for free fields  on either
side of the defect, that the
defect is purely transmitting with the same transmission factor as it would have
had the defect been at rest (with parameter $\eta_0=\eta+\chi$). A more involved calculation
demonstrates that the same is true for the expressions for sine-Gordon
soliton delays calculated in section (2).

\p Returning to the discussion of transmission matrices given in section (3) one might
argue as follows: knowing that the transmission
matrices depend on the rapidity of the scattering soliton only via $\theta-\eta$ when the
defect is stationary, with parameter $\eta$,
suggests that they will be expected (according to \eqref{parameters}) to depend on
$\theta-\eta-\chi=\theta-\eta_0$
when the defect has parameter $\eta$
but moves with rapidity $\chi$. This makes perfect sense, since boosting the
scattering soliton
by $\chi$ would bring the defect with parameter $\eta$ relatively to rest.
The other parameter $\nu$ occurring in \eqref{Tfinalsummary} might then depend upon $\eta$,
but not upon
either $\theta$ or $\chi$.

\p
In the quantum regime it will be necessary to seek an S-matrix for the
defects themselves, call it $^{\rm e}U_{\alpha\beta}^{\gamma\delta}$,
labelled by the initial and final defect topological charges ($\alpha,\beta$)
and ($\gamma,\delta$),
respectively,  satisfying
$\alpha+\beta=\gamma+\delta$, and compatible with the transmission factors.
The defects will be labelled by their parameters and their rapidities so it
is first necessary to decide how their scattering matrix elements might depend on these.
Suppose their parameters if stationary were $\eta_{10}=\eta_1+\chi_1, \
\eta_{20}=\eta_2+\chi_2$, then
the soliton transmission matrix for each defect depends, as argued above, respectively,
on $\theta-\eta_{10}$ and
$\theta-\eta_{20}$. On the other hand, it would be
reasonable to suppose the scattering matrix for
the defects depends  on the difference of their rapidities $\chi_{12}=\chi_1-\chi_2 $,
since whether scattering
takes place at all depends on the sign of the rapidity difference and not on the relative
magnitudes of defect parameters. Besides, a Lorentz transformation effectively shifts
$\theta,\eta_{10},\eta_{20},\chi_1$
and $\chi_2$, leaving $\eta_1, \eta_2$ unchanged.

\p
Using the topological charge labelling, defect-defect scattering will be compatible
with the already determined transmission factors provided
\begin{equation}\label{UTT}
    ^{\rm e}T_{a\alpha}^{b\gamma}(\theta_1,\eta_1)\, ^{\rm e}
    T_{b\beta}^{c\delta}(\theta_2,\eta_2)
    \, ^{\rm e}U_{\gamma\delta}^{\epsilon\rho}(\chi_{12},\eta_1,\eta_2)=\,
    ^{\rm e}U_{\alpha\beta}^{\gamma\delta}(\chi_{12},\eta_1,\eta_2)\,
    ^{\rm e}T_{a\gamma}^{b\epsilon}(\theta_2,\eta_2)\, ^{\rm e}T_{b\delta}^{c\rho}
    (\theta_1 ,\eta_1),
\end{equation}
where $\theta_1=\theta-\eta_{10},\ \theta_2=\theta-\eta_{20}$.
Clearly, $^{\rm e}U$ is infinite dimensional and, as a consequence of the associativity
of \eqref{UTT}, should
itself satisfy a set of Yang-Baxter equations.
Since $\chi_{12}=\eta_{10}-\eta_{20}+(\eta_2-\eta_1)$,
the dependence of $^{\rm e}U$ on the defect parameters could be entirely via the rapidity
difference but that cannot be assumed initially.

\p
Given the conservation of topological charge, it is convenient to define the elements
of $^{\rm e}U$ as follows,
\begin{equation}\label{U}
    ^{\rm e}U_{\alpha\beta}^{\gamma\delta}=\sum_\omega\,
    \sqrt{\nu_1/\nu_2}^{\,\,(\alpha+\beta)/2}\sqrt{\nu_1\nu_2}^{\,\, -\omega/2}\,
    Q^{\omega(\alpha+\beta)/2}\,
    A^\omega_{\alpha\beta}\,
    \delta_\alpha^{\gamma+\omega}\, \delta_\beta^{\delta-\omega},
\end{equation}
and to make use of the symmetrical version of the transmission matrix
given in \eqref{Tfinalsummary}. This expression would be simpler if $\nu_1=\nu_2$,
but this would rule out the possibility of the phase $\nu$ having any dependence on the
defect parameter. On the other hand, using the alternative representation a basis can
always be chosen so that the parameters $\nu_1,\nu_2$ are removed.

\p With this in mind, all the dependence on $\nu_1$ or $\nu_2$ drops out of the
triangle relation and
\eqref{UTT} reduces to a pair of equations ($a=c=+;\ a=+,\ c=-$) which
may be summarized by
\begin{eqnarray}\label{Aequations}
 \nn  A^\omega_{\alpha+2\, \beta} &=& A^\omega_{\alpha\, \beta+2} \\
  Q^{-\omega}p_1 A^\omega_{\alpha\beta}+Q^{-\omega-2}p_2\,A^{\omega+2}_{\alpha\beta} &=&
  Q^\omega p_2A^\omega_{\alpha\, \beta+2} +Q^{\omega+2}p_1\,A^{\omega+2}_{\alpha+2\,\beta},
\end{eqnarray}
where, $$p_k=e^{-\gamma\eta_{k0}},\ k=0,1.$$ The other pair ($a=c=-;\ a=-,\ c=+$)
merely provides the same equations reorganized slightly.

\p Up to now a neat way to write a solution to this has not been found although clearly
the solution depends on the ratio $p_1/p_2=e^{-\gamma(\eta_{10}-\eta_{20})}$.
Since unitarity and crossing will also
impose constraints, the solution to \eqref{Aequations} by itself is not expected to be unique.

\p
The first of \eqref{Aequations} states that $A^\omega_{\alpha\, \beta}$ depends only on
$\alpha+\beta$. Setting $A^\omega_{\alpha\, \beta}=B^\omega_{\alpha+\beta}$, the second of
\eqref{Aequations} becomes
\begin{equation}\label{Bequations}
   \rho^2\, Q^{-\omega}B^\omega_{\kappa}+ Q^{-\omega-2}B^{\omega+2}_{\kappa} =
  Q^{\omega} B^\omega_{\kappa+2} +\rho^2\, Q^{\omega+2}B^{\omega+2}_{\kappa+2},
 \quad  p_1/p_2=\rho^2.
\end{equation}
Defining the generating function
$B(y,z)=\sum_{\omega,\kappa}z^\omega y^\kappa B^\omega_\kappa$, the set of
equations \eqref{Bequations} is
equivalent to the functional relation
\begin{equation}\label{}
B(y,Q^2 z)=y^2\, \left(\frac{1+\rho^2 Q^2 z^2}{\rho^2 + Q^2 z^2}\right)B(y,z).
\end{equation}
The latter can be simplified, formally separating the dependence on $y$ and $z$, by setting
\begin{equation}\label{splitting}
    B(y,z)=e^{(-{i\beta^2}/{4\pi^2})\ln y \ln z}\,  a(y)b(z),
\end{equation}
where $a(y)$ is arbitrary, and
\begin{equation}\label{bfunction}
    b(Q^2 z)=\left(\frac{1+\rho^2 Q^2 z^2}{\rho^2 + Q^2 z^2}\right)\, b(z).
\end{equation}
The function $b(z)$ is assumed to be an analytic function of $z$  in some domain
excluding the origin and infinity. Since $|Q|=1$, if $z$ lies on the unit circle
the multiplier
in \eqref{bfunction} also lies on the unit circle, and therefore
\begin{equation}\label{bmodulus}
    |b(Q^2z)|=|b(z)|.
\end{equation}
If $Q^2$ is a root of unity, iterating \eqref{bfunction} leads eventually to a contradiction
unless $b(z)$ is either zero, or diverges for those values of $Q$. On the other hand,
if $Q^2$ is not a root of unity
there is no such difficulty. One way to solve this problem might be to multiply $b(z)$
by a suitable function of $Q^2$ that possesses a set of zeroes at the roots of unity.
One example might be the Dedekind eta-function with an appropriate argument, i.e. multiply
by a power of
\begin{equation}\label{Dedekind}
    \eta(\tau)= e^{i\pi\tau/12}\prod_{n=1}^\infty\left(1-e^{2i\pi n\tau}\right),\quad
    \tau=\frac{\gamma +1}{2}=\frac{4\pi}{\beta^2},
\end{equation}
and then find a generic solution to \eqref{bfunction}. However, the eta-function is defined
for ${\rm Im}\,\tau>0$ and a small positive imaginary part would have to be given to
$\gamma$. [It is interesting to note that the transformation
$\beta\rightarrow 4\pi i/\beta$ implements the modular transformation
$\tau\rightarrow -1/\tau$.]
The eta-function embodies zeroes at the roots of unity but \eqref{bfunction} gives no
hint as to what power of $\eta(\tau)$ might be appropriate in these circumstances.
A particular solution to \eqref{bfunction} is given by
\begin{equation}\label{bsolution}
    b(z,\chi_{12})=\prod_{k=0}^\infty\, g_k(\eta_{10}-\eta_{20}+\zeta)\,
    g_k(\eta_{10}-\eta_{20}-\zeta),
    \quad z=e^{\gamma\zeta/2},
\end{equation}
where
\begin{equation}\label{g}
    g_k(\lambda)=\frac{\Gamma\left((k+1/2)\tau +1/4+i\gamma\lambda/4\pi\right)\,
    \Gamma\left((k+1/2)\tau +3/4+i\gamma\lambda/4\pi\right)}
    {\Gamma\left((k+1/2)\tau +1/4-i\gamma\lambda/4\pi\right)
    \Gamma\left((k+1/2)\tau +3/4-i\gamma\lambda/4\pi\right)}\, .
\end{equation}
This is straightforward to check using standard properties of the gamma-functions.
Then, any other solution to \eqref{bfunction} may only differ from this by a constant factor,
at least as far as the $z$-dependence is concerned. When $z$ lies on the unit circle it
is easy to check that the
solution given by \eqref{bsolution} and \eqref{g} has unit modulus.

\p An alternative approach would be to label
the defects by pairs of integers, as mentioned in section (3), and to denote their scattering
matrices by
\begin{equation}\label{Upair}
     ^{\rm e}U_{(a,b)\,(b,c)}^{(a,d)\, (d,c)}(\chi_{12},\eta_1,\eta_2)
    \equiv \, ^{\rm e}U_{abc}^{\phantom{a}d\phantom{c}}(\chi_{12}).
\end{equation}
Clearly, only the middle label changes in the scattering process and the second
notation makes use of that fact.

\p
This labelling has an advantage when visualising a two-defect scattering
process or the Yang-Baxter
equations. In effect, the scattering can be represented by a diagram resembling a `quark'
diagram, in which each incoming defect is represented by a pair of lines
labelled by the appropriate integers, $a,b,c,d$, with the incoming defects sharing $b$ and
the outgoing defects sharing $d$. Then the Yang-Baxter equation asserts the equality of
 a pair of diagrams each containing a single closed loop.

\p In this representation,
the triangle relation for $a=c=+$ gives rise to two equations, which may be written:
\begin{equation}\label{}
    ^{\rm e}U_{abc}^{\phantom{a}d\phantom{c}}(\chi_{12})=\,
    ^{\rm e}U_{a+1\,b+1\,c+1}^{\phantom{a+1}d+1\phantom{\,c+1}}(\chi_{12})=
    \,^{\rm e}U_{a-1\,b+1\,c-1}^{\phantom{a-1}d+1\phantom{\,c-1}}(\chi_{12}).
\end{equation}
From these, it is clear that $$^{\rm e}U_{abc}^{\phantom{a}d\phantom{c}}(\chi_{12})=
\,^{\rm e}U_{0\ 0\ c-a}^{\phantom{\ }d-b\phantom{c-a}}(\chi_{12}),$$ where,
in terms of the topological charges, $c-a=\alpha+\beta$ and $b-d=\alpha-\gamma$. Therefore,
the alternative labelling does not appear to provide any more generality than the
labelling in
terms of topological charges.

\p

\p
Apart from \eqref{bfunction}, there are other conditions arising from the
requirements of unitarity and crossing, or the possibility that defects of opposite
charge may annihilate virtually
to the vacuum. In terms of $^{\rm e}U$ the unitarity condition (for $\chi$ real) is,
\begin{equation}\label{Uunitarity}
^{\rm e}U_{\alpha\beta}^{\gamma\delta}(\chi) \,
^{\rm e}\bar U_{\delta\gamma}^{\epsilon\rho}(\chi)
=\delta_\alpha^{\rho}\,\delta_\beta^\epsilon,
\end{equation}
and this translates in terms of $B^\omega_\kappa$ into the collection of relations,
\begin{eqnarray}
 & & \sum_\omega\, B_\kappa^\omega\, \bar B^{\omega+\epsilon}_\kappa =\delta_0^\epsilon,
\end{eqnarray}
which ought to hold for each choice of $\epsilon$ and $\kappa$. This may be rewritten
in terms of generating functions setting $\bar B(y,z)=\sum_{\omega\, \kappa}
z^{-\omega}\, y^{-\kappa}\,\bar B_\kappa^\omega$. If both $y$ and $z$ lie on unit
circles, $\bar B(y,z)$ is the complex conjugate of $B(y,z)$. Then, using the separation
property of $B(y,z)$ given in \eqref{splitting}, the unitarity condition itself
separates into two types of term
\begin{equation}\label{zcondition}
    \oint\frac{dz}{2\pi i z}z^\epsilon b(z) \bar b (z) =0,\ \hbox{for}\ \epsilon \ne 0,
\end{equation}
and
\begin{equation}\label{ycondition}
     \oint\frac{dz}{2\pi i z} b(z) \bar b (z)\,\oint\frac{dy}{2\pi i y}y^{-\kappa}
     a(y)\, \oint\frac{dy^\prime}{2\pi i y^\prime}y^{\prime \kappa}\,
      \bar a(y^\prime)=1,\ \hbox{for\
     each}\ \kappa.
\end{equation}
In all cases it has been assumed there is a singularity-free region enclosing the origin
and containing the contours of integration. Equation \eqref{zcondition}
follows automatically on
choosing the $z$ contour to lie on the unit circle since it has been noted already that
$b(z)$ has unit modulus there. The other part of the condition is more problematical
since it appears to require that each coefficient in a Laurent expansion of $a(y)$
is equal to the inverse of the corresponding coefficient in a similar expansion of
$\bar a (y)$, at least  up to an overall, term-independent constant.

\p It is tempting to suppose the crossing property should take the form
\begin{equation}\label{Ucrossing}
    U_{\alpha\, \beta}^{\gamma\, \delta}(i\pi-\chi_{12},\eta_1,\eta_2)=
    U_{\alpha\,\,-\gamma}^{-\beta\,\, \delta}(\chi_{12},\eta_1,\eta_2),
\end{equation}
which, if $\nu_1=\nu_2=\nu$ translates in terms of $B^\omega_\kappa$ to,
\begin{equation}\label{Bcrossing}
    B_\kappa^\omega(i\pi-\chi_{12})= \nu^{(\omega-\kappa)/2}\,B^\kappa_\omega(\chi_{12}).
\end{equation}
Because the upper and lower indices on $B_\kappa^\omega$ are interchanged
it will be possible to use the crossing relation to relate the two functions
$a(y)$ and $b(z)$. Rewriting
\eqref{Bcrossing} in terms of the generating functions leads to the relation,
\begin{equation}\label{abrelation}
    \frac{a(y, i\pi-\chi_{12})}{b(y/\sqrt{\nu},\chi_{12})}\, e^{(i\beta^2/8\pi^2)\ln\nu\,\ln y}
    =\frac{a(\sqrt{\nu}\,z,\chi_{12})}{b(z,i\pi-\chi_{12})}\,
    e^{(i\beta^2/8\pi^2)\ln\nu\,\ln \sqrt{\nu}\,z},
\end{equation}
from which it is clear both sides must be independent of $y$ and $z$, and separately
equal to a crossing symmetric function $c(\chi_{12})=c(i\pi-\chi_{12})$. Then, either side
of \eqref{abrelation} implies an expression for the  unkown function $a(y)$ in terms of
$b(y)$, itself given by \eqref{bsolution}. Explicitly, the expression for $a(y)$ is,
\begin{equation}\label{aexpression}
    a(y,\chi_{12})=c(\chi_{12})\, b(y/\sqrt{\nu},\, i\pi-\chi_{12})\,
    e^{-(i\beta^2/8\pi^2)\ln\nu\,\ln y}.
\end{equation}
Clearly, the expression \eqref{aexpression} simplifies significantly if $\nu=1$.

\p
Alternatively, the analogue of \eqref{annihilation} representing the possibility for a
defect to annihilate virtually with a defect of opposite topological charge would be
\begin{equation}\label{Uannihilation}
    c^0_{\beta\, -\beta}\, \delta_\alpha^\rho=\sum_{\epsilon,\,\gamma}\, ^{\rm e}
    U_{\alpha\beta}^{\epsilon\gamma}(\chi-i\pi/2)\,
^{\rm e}U_{\gamma \, -\beta}^{-\epsilon\,\rho}(\chi+i\pi/2)\, c^0_{\epsilon\, -\epsilon},
\end{equation}
and in terms of $B^\omega_\kappa$ this becomes
\begin{equation}\label{Bannihilation}
    c^0_{\beta\, -\beta}=\sum_\epsilon\, Q^{\epsilon\alpha}\, \nu^{-(\epsilon+\alpha)/2}\,
    B^\epsilon_\alpha(\chi-i\pi/2)\, B_\epsilon^\alpha(\chi+i\pi/2)\,
    c^0_{\alpha-\beta-\epsilon\,
    \epsilon+\beta-\alpha}\, .
\end{equation}
However, without detailed knowledge of the couplings it is not easy to extract
information from this
relationship.

\section{Discussion}
\jumpup
The jump-defect possesses a variety of interesting classical properties stemming
 from the fact that the integrable defect conditions are a `frozen' B\"acklund transformation.
 The purpose of this article has been to explore the corresponding quantum sine-Gordon field
 theory and investigate the extent to which the classical picture extends to the quantum
 domain. One of the striking features of the classical scattering of a soliton with a
 jump-defect was the possibility of the soliton being `eaten' by the defect, and another was
 the possibility of the soliton converting to an anti-soliton. The fact a soliton
 can disappear
 and be replaced by an `excited' defect, carrying exactly the energy and momentum of the
 soliton, emerges in the quantum field theory as a resonant soliton-defect bound state,
 with a finite, coupling-dependent width. Scattering with the excited state is not expected
 to be unitary (since the excited defect is no longer a possible asymptotic state)
 and that turns out to be precisely the case. Curiously, the defect treats solitons
 and anti-solitons
 differently. The algebraic analysis of the triangle relations makes this clear. Morever,
 an argument presented in section (3), based on a functional integral representation of the
 transmission matrix,
 demonstrates how even this fact follows from the Lagrangian starting point \eqref{sGdefect}.
 Another surprising feature appears when
 one considers the
 breather transmission factors. Each of these may be defined via a bootstrap procedure.
 However,  the transmission factor for the lightest breather is
 predicted to be completely independent of the coupling constant. This fact, if correct,
 should be verifiable in perturbation theory by calculating the quantum corrections
 to the classical result given in section (2).
 On the other hand, if the prediction
 is false, it would indicate the necessity of additional CDD factors in the derived
 transmission matrix presented in \eqref{Tfinalsummary}, and a detailed perturbative
 calculation may
 provide a pointer towards the precise form of any such factors.
 Typically, the other breather transmission factors
 have poles at complex rapidity whose interpretation is obscure owing
 to the lack of classical configurations that might correspond to them. Possibly they
 indicate a collection of new excited states of the
 even-charged defect whose details will depend upon the value of the bulk coupling.
 Or, they may be artifacts of the bootstrap procedure. In either case, a detailed
 explanation is needed.

 \p There has been work previously on models containing unstable bound states
 \cite{Miramontes},
 including interesting ideas concerning their consequences in the context of impurities
 \cite{Fring03}. However, it
 appears the present work may provide the first example of unstable states appearing naturally
 within the sine-Gordon model itself. For the reasons stated, it is highly desirable to
 find a physical situation where the sine-Gordon B\"acklund transformation is part of
 the formulation. If it exists a physical system of
 this kind would offer the opportunity to control
 solitons and have possible consequences for communications making use of solitons.

 \p
 There is no reason why defects should not be able to move with constant speed, and
 therefore to scatter. Classically, the delays, or changes of character, suffered by a
 soliton scattering with two moving defects is entirely consistent once the defects
 have changed places. In fact the consistency is an expression of Bianchi's identity
 concerning the permutability of a pair of B\"acklund transformations. The classical
 picture of a moving defect is analyzed in detail and it has been verified that with
 a natural choice of defect condition, the jump-defect moves at constant speed. Moreover,
 multiple defects have no long range interactions with each other. For this reason
 it is plausible that their interaction in the quantum domain is local and factorisable,
 implying that there should be  a triangle compatibility  relation \eqref{UTT},
 involving the scattering matrix
 for a pair of defects and the already derived transmission matrices for the solitons.
 The triangle relation is supplemented by unitarity and crossing
 relations and a fuller discussion
 will be given in a subsequent paper. In this regard, the purpose of this
 article has been to develop
 the framework within which such questions may be asked and to give an indication
 of how they may be tackled. Further investigation will be needed to elaborate all the details.

 \p
 Since the sine-Gordon model allows both integrable boundary conditions and integrable
 jump-defects one might also ask how defects behave when they encounter a boundary.
 To some extent, the question should be answerable  algebraically by finding solutions to
 the associated reflection Yang-Baxter equations \cite{Cherednik}, once the defect
 scattering matrix is completely determined - although such solutions would need to be
 supplemented by  other information. One might speculate that placing a
 stationary defect close to a boundary could modify the boundary condition.
 The jump-defects provide an environment where such questions
 can be asked, and answered, up to a point. In the $\sinh$-Gordon model
 all the integrable boundary conditions and reflection factors are known
 \cite{Ghoshal}. However,
 the scalar particle has a $\beta$-independent transmission factor that would appear to be
 quite unable
 to account for the intricate $\beta$-dependence of the reflection factors for general
 boundary conditions, starting
 from, for example, the reflection factor associated with a Neumann
 boundary condition. That is,
of course, presuming the results
 of section (5) are entirely unmodified by CDD factors. Otherwise, it might be possible to
 invert the argument and use known reflection factors to inform the discussion
 of transmission factors.

\p
 A more interesting question for the future will be to decide whether
 the jump-defects themselves might be described by a local quantum field theory coupled to
 (or, even a part of) the sine-Gordon model. This was hinted at in the introduction
 where it was pointed out that a pair of stationary, classical, jump-defects behave
 like a soliton, at least in so far as mimicking the delay a soliton would experience
 overtaking another soliton. It might be fruitful to regard solitons
 as bound pairs of defects. In which case, whatever the theory of defects is,
 it should contain
 the sine-Gordon model, perhaps as a suitable limit. One might also wonder about the
 dual relationship
 between the
 sine-Gordon model and the massive Thirring model, formulated by Skyrme, Coleman and
 Mandelstam \cite{Mandelstam}. If the jump-defects have a role to play within the sine-Gordon
 model, then
 they should also be describable within the massive Thirring model where the basic fields
  are fermions equivalent to the solitons of the sine-Gordon model. One might imagine a hierarchy in
 which the defects bind to form solitons, and the solitons bind to form breathers.
 This  aspect needs  further investigation.

 \p There are other integrable models with classical soliton behaviour and it will be
 interesting to see if the jump-defect idea may be incorporated naturally within those models.
 Some results are already known for the affine Toda field theories, especially those of
 $a_n$ type \cite{bcztoda}. However, the results known so far are restricted to classical
 integrability, and placing the jump-defects in a quantum context for these examples
  has yet to be accomplished.

\vskip 1.0cm
\noindent{\bf Acknowledgements}
\vskip .25cm
\noindent
CZ thanks JSPS for a Fellowship and EC is indebted to members
of the Ecole Normale Superieure de Lyon, the University of Bologna and the Yukawa Institute
for Theoretical Physics, and especially Jean Michel Maillet, Francesco
Ravanini and Ryu Sasaki, for their hospitality. PB thanks Patrick Dorey for discussions and
EC wishes to thank Davide Fioravanti for comments.
The work has also been supported by the Leverhulme Trust and
EUCLID - a European Commission RTN Network (contract
HPRN-CT-2002-00325). EC would like to thank the organisers of
the 2nd EUCLID Annual Conference (September 2004, Sozopol, Bulgaria) and PB thanks
the organisers of the 3rd EUCLID Spring School (May 2005, Trieste, Italy) for the
opportunity to present some of the ideas discussed in this article.

\end{document}